\documentclass[nonblindrev,copyedit]{informs3}    % spaced for copyediting
\usepackage{natbib}
 \bibpunct[, ]{(}{)}{,}{a}{}{,}%
 \def\bibfont{\small}%
\usepackage{multirow}
\usepackage{slashbox}
\usepackage{multirow}
\usepackage{rotating}
\usepackage{colortbl}
\usepackage{setspace}

\usepackage{graphicx}
\usepackage{graphics}
\usepackage{Input}
\usepackage{qtree}
\usepackage{amsmath,amsfonts,latexsym,amssymb}
\usepackage{ifthen}
\usepackage[gen]{eurosym}
\usepackage{colortbl}

\usepackage{wasysym}
\usepackage[vlined, ruled]{algorithm2e}

\newboolean{includeHidden}
\setboolean{includeHidden}{false}
\usepackage[]{todonotes}
\newcommand{\hide}[1]{\ifthenelse{\boolean{includeHidden}}{{\tiny\textbf{HIDDEN:~}#1}}{}}

\DeclareMathOperator*{\Min}{\mathop{Min}\limits_{\mathrm{s.t.}}}

\newcommand{\set}[2]{\left\{\,#1\mid #2\,\right\}} %Menge mit Mittelsteg
\providecommand{\abs}[1]{\left\lvert #1 \right\rvert} %Betrag
\renewcommand{\min}[1]{\mathop{min}\,\left\{\,#1\,\right\}} %max/min Menge

\newcommand{\mins}[2]{\mathop{min}\,\set{#1}{#2}} %max/min mit Set

\newcommand{\rank}[2]{rank\rbr{#1, #2}}

\newcommand{\sdpref}[1]{\succ_{#1}^{SD}}
\newcommand{\sdprefeq}[1]{\succeq_{#1}^{SD}}

\newcommand{\rbr}[1]{ \left( #1 \right) }	%round bracket
\newcommand{\trans}{^\tau}	

\usepackage{tikz}
\usetikzlibrary{calc,fit}
\usetikzlibrary{positioning,chains}

\makeatletter

\def\tmk@labeldef#1,#2\@nil{%
  \def\tmk@label{#1}%
  \def\tmk@def{#2}%
}

\makeatother

\include{hyphenation}

\TheoremsNumberedThrough     % Preferred (Theorem 1, Lemma 1, Theorem 2)
\EquationsNumberedThrough    % Default: (1), (2), ...
\MANUSCRIPTNO{0001}

\begin{document}
%\setcounter{page}{1}
%\DOI{10.1287/xxxx.0000.0000}%

\RUNAUTHOR{Bichler, Merting, and Uzunoglu}

%\RUNTITLE{Coordination Mechanisms for Retail Transportation Logistics}
\RUNTITLE{Assigning Course Schedules}

%\TITLE{Coordination Mechanisms for Retail Transportation Logistics}
\TITLE{Assigning Course Schedules: About Preference Elicitation, Fairness, and Truthfulness}

\ARTICLEAUTHORS{%
\AUTHOR{Martin Bichler, Soeren Merting}
\AFF{Department of Informatics, 85748 Garching, Technical University of Munich, Germany, \{bichler, soeren.merting\}@in.tum.de}
\AUTHOR{Aykut Uzunoglu}
\AFF{Faculty of Business and Economics, 86135 Augsburg, University of Augsburg, Germany, aykut.uzunoglu@wiwi.uni-augsburg.de}

} % end of the block

\ABSTRACT{%
Course assignment is a wide-spread problem in education and beyond. Often students have preferences for bundles of course seats or course schedules over the week, which need to be considered. The problem is a challenging distributed scheduling task requiring decision support. First-Come First-Served (FCFS) is simple and the most widely used assignment rule in practice, but it leads to inefficient outcomes and envy in the allocation. Recent theoretical results suggest alternatives with attractive economic and computational properties. Bundled Probabilistic Serial (BPS) is a randomized mechanism satisfying ordinal efficiency, envy-freeness, and weak strategy-proofness. This mechanism also runs in polynomial time, which is important for the large problem instances in the field. We report empirical results from a first implementation of BPS at the Technical University of Munich, which allows us to provide important empirical metrics such as the size of the resulting matching, the average rank, the profile, and the popularity of the assignments. These metrics were central for the adoption of BPS. In particular, we compare these metrics to Random Serial Dictatorship with bundle bids (BRSD). The BRSD mechanism is used to simulate the wide-spread First-Come First-Served (FCFS) mechanism and it allows us to compare FCFS (BRSD) and BPS. While theoretically appealing, preference elicitation is a major challenge when considering preferences over exponentially many packages. We introduce tools to elicit preferences which reduce the number of parameters a student needs to a manageable set. The approach together with BPS yields a computationally effective tool to solve course assignment problems with thousands of students, and possibly provides an approach for other distributed scheduling tasks in organizations. 
}

%carrier coordination, approximation mechanism, combinatorial auction, market design, time slot management}

\maketitle
{\begin{center}Working Paper: Version \today \end{center}}

\section{Introduction}

Course assignment is arguably one of the most wide-spread assignment problems where money cannot be used to allocate scarce resources. Such problems appear at most educational institutions. Matching with preferences has received significant attention in the recent years. While simple first-come first-served (FCFS) rules are still wide-spread, many organizations adopted matching mechanisms such as the deferred acceptance algorithm \citep{Gale1962, diebold2014course} or course bidding \citep{Sonmez10, krishna2008research} to allocate scarce course seats. Although many course assignment problems are similar to the widely studied school choice problems with students private preferences for one out of many courses, other applications differ significantly. In particular, students are often interested in schedules of courses across the week. Assigning schedules of courses has been referred to as the \textit{combinatorial assignment problem} (CAP) \citep{Budish2011}. Similar problems arise when siblings should be assigned to the same schools in school choice \citep{Abdulkadiroglu-et-al-2006}, or couples in the context of the hospital residency matching \citep{ashlagi2014stability}. Overall, the CAP can be seen as a general form of a distributed scheduling problem.

Although there is a huge body of literature on scheduling, the CAP is specific in a number of ways. First, we can only elicit ordinal preferences and no money must exchange hands. Second, students have private preferences over course schedules and we want to have mechanisms that incentivize students to reveal these preferences truthfully. Third, apart from efficiency, fairness of the allocation is an important concern in matching with preferences \citep{Roth1982}. Fourth, the allocation of course schedules is a computationally hard (NP-hard) problem and for the problem sizes with hundreds of students an exact solution might not be tractable.

The need to assign course schedules rather than courses individually became apparent in an application of matching with preferences at the Technical University of Munich that we will discuss. The Department of Informatics is using the deferred acceptance algorithm for two-sided matching problems and random serial dictatorship for one-sided matching problems. These algorithms are used to assign seminars or practical courses, and every semester about 1500 students are being matched centrally \citep{diebold2014course}. For seminars and practical courses students need to get assigned one out of many courses offered per semester.

In the initial three semesters the situation is different. There are large courses with hundreds of students (e.g. on linear algebra or algorithms). These courses include a lecture and small tutor groups. Students need to attend one tutor group for three to four courses in each semester. These tutor groups should not overlap and they should be adjacent to each other such that students do not have a long commute for each of the tutor groups individually. For example, a student might want to have two tutorials in the morning and one after lunch on a particular day to reduce his commute time, and he would have a strong preference for this schedule over one where the tutorials are scattered across the week. In any case, students have timely preferences over course schedules that need to be considered, which makes it a combinatorial assignment problem. These problems are wide-spread in academia.

A first and seminal approach to address this challenging problem, the \emph{approximate competitive equilibrium from equal incomes mechanism }(A-CEEI), was published by \cite{Budish2011}. 
In A-CEEI students report their complete preferences over schedules of courses, the mechanism assigns a budget of fake money to each student that she can use to purchase packages (or schedules) of courses. Then an optimization-based mechanism computes approximate competitive equilibrium prices, and the student is allocated her most preferred bundle given the preferences, budgets, and prices. 
It is well known that serial dictatorships are the only strategy-proof and efficient mechanisms for multi-unit and also combinatorial assignment problems \citep{Papai01, Ehlers03}. A-CEEI is relaxing design goals such as strategy-proofness and envy-freeness to approximate notions, which makes it a remarkable and practical contribution to a fundamentally hard problem. 
The mechanism has been shown to be approximately strategy-proof, approximately envy-free, and Pareto efficient. 
\cite{Budish2017} reports the empirical results at the Wharton School of Business. In addition, \cite{Budish2018} summarize the results of lab experiments. 

The work was breaking new ground, but the A-CEEI mechanism is also challenging. First, it is not guaranteed that a price vector and course allocation exists that satisfies all capacity constraints. This is not surprising given that prices are linear and anonymous. Second, the problem of computing the allocation problem in A-CEEI is PPAD-complete and the algorithms proposed might not scale to larger problem sizes required in the field \citep{othman2016complexity}. Third, students might not be able to rank-order an exponential set of bundles, which is a well-known problem (aka. missing bids problem) in the literature on combinatorial auctions (with money) \citep{Milgrom09, Bichler11a, Bichler14}. The latter is a general problem in CAP not restricted to A-CEEI, which we will discuss in much more detail below.

Randomization can be a powerful tool in the design of algorithms, but also in the design of economic mechanisms. \cite{nguyen2015assignment} recently provided two randomized mechanisms for one-sided matching problems, one with cardinal and one with ordinal preferences for bundles of objects. The mechanism for ordinal preferences is a generalization of probabilistic serial \citep{bogomolnaia2001new}, called Bundled Probabilistic Serial~(BPS). \cite{nguyen2015assignment} show that this randomized mechanism is ordinally efficient, envy-free, and weakly strategy-proof. These appealing properties come at the expense of feasibility, but the constraint violations are limited by the size of the packages. In course assignment problems the size of the packages is typically small (e.g., packages with three to four tutor groups) compared to the capacity of the courses or tutor groups (around 30 seats or more). There is no need for prices or budgets, and computationally the mechanism runs in polynomial time, which is important for large instances of the course allocation problem that can frequently be found. This makes BPS a practical approach to many problems that appear in practice.

\subsection{Contributions}

We report on a first large-scale field study of BPS and address important problems in the implementation of mechanisms for the combinatorial assignment problem that are beyond a purely theoretical treatment. 
In particular, preference elicitation is a central concern in combinatorial mechanisms with a fully expressive bid language and we provide a practical approach that addresses the combinatorial explosion of possible packages for many applications. 
Theoretical contributions of assignment mechanisms largely focus on envy-freeness and efficiency as primary design desiderata. We report properties of matchings such as their size, their average rank, the probability of matching, the profile, and the popularity. These properties are of central importance for the choice of mechanisms. For IS designers it is important to understand the trade-offs with other mechanism, in particular with the wide-spread FCFS.

%We compare BPS to First-Come First-Served~(FCFS), which is the de facto standard at most universities. 

Implementing and testing new IS artifacts for coordination in organizations is challenging and we are grateful for the possibility to run a large-scale field experiment at the Department of Informatics of the Technical University of Munich (TUM). This is particularly true for a non-trivial mechanism such as BPS, which involves advanced optimization and randomization. Yet, we can report on the assignment of 1439 students in the summer term 2017 to 67 tutor groups for 4 classes and the assignment of 1778 students in the winter term 2017/2018 to 66 tutor groups for 4 classes using BPS.\footnote{Not all students submitted a non empty  preference list. Therefore, we consider in our evaluation only 1415 students in summer term and 1736 students in winter term.} Based on this data the department has adopted the new mechanism for good.

For such a large application we could not elicit preferences of students for BPS and let them participate in FCFS simultaneously. Instead we simulated FCFS via a version of Random Serial Dictatorship that allows for bundles (BRSD), which is of independent interest as an assignment mechanism. In our numerical experiments we simulated FCFS via a large number of random order arrivals in BRSD using the preferences elicited in BPS and average across all of them. This approach allows for a comparison between BPS and BRSD (FCFS) on equal footing. 
%This is because a part of the students participating in BPS reported in our survey that they were not bidding truthfully. However, the comparison gives us a reasonable estimate for the differences between the two mechanisms. Note that strategic bidding is no an issue in BRSD which is obviously strategy-proof as we will discuss.

FCFS only collects limited information about the preferences of participants, a single package only. Mechanisms for the combinatorial assignment problem allow participants to specify preferences for all possible packages. However, a fully enumerative bid language requires participants to submit preferences for an exponential set of packages which is impractical. 
%Sometimes the application domain suggests value functions which only require the elicitation of a few parameters. Compact bid languages have been applied successfully in procurement and spectrum auction design \citep{Bichler11a, Bichler14}, and they allow bidders to describe their preferences with a low number of parameters. \cite{Budish2017} use a course-level scoring, but write that ``optimal language design is an interesting open question for future research.'' Indeed, a course-level scoring rule would not be expressive enough to describe the ranking of bundles that we find in our course scheduling application. 
Preference elicitation and user interface design have long been a topic in IS research \citep{SantosMS88, Benbasat2011}. We contribute an approach that is applicable in a wide array of CAP applications where timely preferences matter. We elicit a small number of parameters about breaks and preferred times and days of the week. Together with some prior knowledge about student preferences this allows us to score and rank-order all possible packages. Students could iteratively adapt the parameters and the ranking, which then served as an input for BPS. While such ranking algorithms will differ among types of applications, adequate decision support that aids the ranking of exponentially many packages is a crucial prerequisite to actually achieve the benefits of combinatorial assignment in real-world applications. 

In our empirical analysis, we show that BPS has many advantages over BRSD in all of the properties introduced earlier. While the differences in these criteria are small, envy-freeness turns out to be the most compelling advantage of BPS. The level of envy that we find in BRSD is substantial in spite of the limited complementarities in student preferences, who are only interested in packages with at most four tutor groups. This has to be traded off with the simplicity of FCFS. Overall, we \textit{empirically test and illustrate theory} that has been developed only recently. We show that randomized matching mechanisms together with appropriate decision support tools are a powerful new IS design recipe for daunting coordination problems in organizations. Thus, we contribute to the traditional IS research stream in decision support and design science research, but introduce new methods and applications \citep{banker200450th}.
%Actually, a part of the students indicated strategic reporting (either hiding the most preferred or least preferred time slots) in our survey. Our results are based on the assumption that we have a complete ranking of students for all possible packages that reflects their true preferences. This is all but trivial, and the advantages of BPS depend crucially on the way how preferences are elicited. Simple course-level scoring can be insufficient, as we show.

\section{Combinatorial Assignment Problems}

Let us now define the combinatorial assignment problem (CAP) in the context of course assignment applications, desirable properties, and randomized mechanisms.

\subsection{Assignment Problems} 

Assigning objects to agents with preferences but without money is a fundamental problem referred to as \emph{assignment problem with preferences} or \emph{one-sided matching with preferences}. We will use the term assignment or matching interchangeably. In course assignment, students express ordinal preferences which need to be considered in the assignment. A \emph{one-sided one-to-many course assignment problem} consists of a finite set of $n$ students (or agents) $S$ and a finite set of $m$ courses (or objects) $C$ with the \emph{maximum capacities} $q = (q_{{1}}, q_{{2}}, \dots, q_{{m}})$. 

In the \emph{combinatorial assignment problem} in the context of course allocation, every student $i \in S$ has a complete and transitive preference relation $\succeq_i \in \mathcal{P}$ over subsets (or bundles) of elements of $C$. 
A preference profile $\succeq = (\succeq_1, \cdots, \succeq_n)\in \mathcal{P}^{|S|}$ is an $n$-tuple of preference relations. 
%The \emph{preferences} of the students ($\succeq$) do not have to be strict in general, they could also contain \emph{indifferences (ties)}. However, ties make the problem much harder as we will discuss in the conclusions. 
For most of the paper we will assume strict preferences, but we discuss indifferences in the conclusions.  
We can model the demand of the students with binary vectors $b \in \{0,1\}^{m}$, where $b_{j} = 1$ if course $j$ is included in $b$.  We define the size of a bundle $b$ with $size(b) = \sum_{j=1}^{m}b_j$, the number of different courses included in the bundle. Let $B$ be the set of all feasible bundles $b$. Let $x_{ib}$ be a binary variable describing if bundle $b$ is assigned to student $i$. Then we can model the demand and supply as linear constraints. The supply constraints make sure that the capacity of the courses are not exceeded, and the demand constraints determine that each student can win at most one bundle. 

\begin{subequations}
\begin{align} 
	& \sum\limits_{i\in I, b \in B} x_{ib}b_j \leq q_j & j \in C \label{eq:supply}\tag{supply}\\
	& \sum\limits_{b \in B} x_{ib} \leq 1 & \forall i \in S \label{eq:demand}\tag{demand}\\
  	& x_{ib} \in \{0,1\} & \forall i \in S, b \in B \label{eq:binary}\tag{binary} 
\end{align}
\end{subequations}

Courses in our application are actually tutor groups and each tutor group belongs to one of $\ell$ classes. Students in our application can only select bundles with at most one tutor group in each of these classes. For example, a student might select a bundle with a course seat in a tutor group for mathematics on Monday at 1 pm, and another tutor group in software engineering two hours later, but no additional tutor group in mathematics or software engineering in this bundle. As a result, the possible size of a bundle $b$ is $size(b) \leq \ell \ll m$. The Web interface takes care that students only submit valid bundles, which have at most one tutor group for each of the $\ell$ classes and a size less than or equal to $\ell$.

A \emph{deterministic combinatorial assignment} (deterministic matching) is a mapping $M \subset S \times B$ of students $S$ and bundles $B$  of courses $C$. $\mathcal{M}$ describes the set of all deterministic matchings. A matching is \emph{feasible} if it is a feasible integer solution to the constraints \ref{eq:demand} and \ref{eq:supply}. 
\emph{Random combinatorial assignments} (random matchings) are related to fractional assignments with $0 \leq x_{ib} \leq 1$ and random assignment mechanisms can be used to fractionally allocate bundles of course seats to students.  

For (non-combinatorial) assignment problems with single-unit demands the Birkhoff-von-Neumann theorem \citep{birkhoff1946three,von1953certain} says that every fractional allocation can be written as a unique probability distribution over feasible deterministic assignments.
That is, any random assignment can be implemented as a lottery over feasible deterministic assignments, such that the expected outcome of this lottery equals the random assignment. 
One can describe a random assignment as a bistochastic matrix, where $p_{ic}$ is the probability that student $i$ is assigned to course $c$. The Birkhoff-von-Neumann theorem shows that such a bistochastic matrix can be decomposed into a convex combination of permutation matrices, which describe feasible deterministic assignments. However, the Birkhoff-von-Neumann theorem fails when bundles of course seats need to be assigned.
\cite{nguyen2015assignment} generalize this result and show that any fractional solution respecting the \ref{eq:demand} and \ref{eq:supply} constraints can be implemented as a lottery over integral allocations that violate the \ref{eq:supply} constraints only by at most $\ell-1$ course seats. 
%The fractional solution $x_{ib}$ to the \ref{eq:demand} and \ref{eq:supply} constraints is then equal to the probability that agent $i$ obtains bundle $b$.

\subsection{Design Desiderata}
\label{sec:designdes}

Efficiency, envy-freeness, and strategy-proofness are design desiderata of first-order importance typically considered in the theoretical literature on deterministic assignment problems. For randomized mechanisms one has to reconsider these design desiderata and we will briefly introduce relevant definitions in this section. Stochastic dominance (SD) is the key concept among all of these definitions as it provides a natural way to compare random assignments. Let $\Delta$ describe the set of all possible random matchings. With $p_i$ we refer to the assignment of student $i$ in the random matching $p$, and denote with $p_{ib}$ the probability that student $i$ gets allocated bundle $b$. We will omit the subscript $i$ when it is clear which student is meant. 
Given two random assignments $p, q \in \Delta$, student $i$ \emph{$SD$-prefers} $p$ to $q$ if, for every bundle $b$, the probability that $p$ yields a bundle at least as good as $b$ is at least as large as the probability that $q$ yields a bundle at least as good as $b$. 

\begin{definition}[SD-prefer]
A student $i\in S$ \emph{$SD$-prefers} an assignment $p\in \Delta$ over $q\in \Delta$, $p \sdprefeq{i} q$, if 
\begin{equation}
\sum_{b' \succeq_i b} p_{ib'} \geq \sum_{b' \succeq_i b} q_{ib'},   \forall b \in B
\end{equation}
\end{definition}

In other words, a student $i$ prefers the random assignment $p$ to the random assignment $q$ if $p_i$ stochastically dominates $q_i$. Note, that $\sdprefeq{}$ is not a complete relation. That is there might be assignments $p$ and $q$, which are not comparable with this relation.
First-order stochastic dominance holds for all increasing utility functions and implies second-order stochastic dominance, which is defined on increasing concave (risk-averse) utility functions. In other words, risk-averse expected-utility maximizers prefer a second-order stochastically dominant gamble to a dominated one.

\cite{nguyen2015assignment} show that a lottery over allocations of bundles induces probability shares over these bundles that satisfy \ref{eq:demand} and \ref{eq:supply} constraints. Thus a lottery coincides with a fractional solution to both constraints. However, a fractional solution respecting \ref{eq:demand} and \ref{eq:supply} does not need to have a lottery over deterministic assignments.

%Such a fractional assignment $x$ weakly stochastically dominates an allocation $y$ for agent $i$, if  $\sum_{b' \succeq_i b} x_{ib'} \geq \sum_{b' \succeq_i b} y_{ib'}$ for all $b \subseteq B$. The lottery $x$ stochastically dominates $y$, if $x$ weakly stochastically dominates $y$ and this equation is strict for some bundle $b$.  

%\subsubsection{Efficiency}
One desirable property of matchings is \emph{(Pareto) efficiency} such that no student can be made better off without making any other student worse off. A \emph{deterministic} matching $M$ is \emph{efficient} with respect to the students if there is no other feasible matching $M'$ such that $M'(i) \succeq_{i} M(i)$ for all students $i \in S$ and $M'(i) \succ_{i} M(i)$  for some $i \in S$. One can generalize this to random matchings and lotteries:

\begin{definition}[Efficiency]
A random assignment $p\in \Delta$ is \emph{ex post efficient}, if $p$ can be implemented into a lottery over Pareto efficient deterministic assignments. A random assignment $p\in \Delta$ is \emph{ordinally efficient}, if there exists no random assignment $q$ such that $q$ stochastically dominates $p$, i.e. $\nexists q\in \Delta: \forall i\in S: q \sdprefeq{i} p$ and $\exists i\in S: q \sdpref{i} p$. 
\end{definition}

%A \textit{random} assignment is \textit{ex post efficient} if it can be represented as a probability distribution over Pareto efficient deterministic assignments. 
Ordinal efficiency comes from the Pareto ordering induced by the stochastic dominance relations of individual students.
%A random assignment $p$ is \textit{ordinally efficient} if there exists no random assignment $q$ such that $q$ stochastically dominates $p$, i.e. $p_i \succeq_i^{SD} q_i$ for all $i \in S$ and $p_i \succ_i^{SD} q_i$ for some $i \in S$. 
It can be shown that ordinal efficiency implies ex post efficiency \citep{bogomolnaia2001new}.

%\subsubsection{Fairness}		
\emph{Fairness} is another important design goal. A basic notion of fairness for randomized assignments is the \emph{equal treatment of equals}, i.e. students with identical preferences receive identical (symmetric) random allocations. Envy-freeness is stronger.
% In the absence of money, a student could claim a large utility for its most preferred bundle of course seats. 

\begin{definition}[Envy-Freeness]\label{def:envy}
A random assignment $p\in \Delta$ is \emph{(strongly) $SD$-envy-free}, if $\forall i,j\in S: p_i \sdprefeq{i} p_j$. We call $p$ \emph{weakly $SD$-envy-free}, if $\nexists i,j\in S: p_j \sdpref{i} p_i$.
\end{definition}

%$SD$-envy-freeness is a desirable property in these situations.%
$SD$-envy-freeness means that student $i$ weakly $SD$-prefers the random matching she is faced with to the random assignment offered to any other student, i.e., a student's allocation stochastically dominates the outcome of every other student. For weak $SD$-envy freeness it is only demanded that no student's allocation is stochastically dominated by the allocation of another student. $SD$-envy-freeness implies equal treatment of equals.
%We will omit $SD$ as a prefix for brevity.

%An assignment mechanism is an algorithm, which computes a matching $\mu$ for given preferences of students. 
%More formally, a deterministic \emph{one-sided matching mechanism} $\chi$ is a function $\chi : \mathcal{P}^{|S|} \rightarrow 2^{m}$ that returns a feasible matching of students to courses for every preference profile of the students. 
A \emph{randomized assignment mechanism} is a function $\psi : \mathcal{P}^{|S|} \rightarrow \Delta$ that returns a random matching $p \in \Delta$. The mechanism $\psi(\succeq)=p$ is \emph{ordinally efficient} if it produces ordinally efficient allocations.
%We call $\psi$ \emph{ex post Pareto efficient}, if $p$ is a convex combination\index{convex combination} of Pareto optimal matchings. 
In terms of fairness, one could aim for a matching where equals are treated equally. We call a randomized matching mechanism $\psi$ \emph{symmetric}, if for every pair of students $i$ and $j$ with $\succeq_{i}=\succeq_{j}$ also $p_{i}=p_{j}$. This means that students who have the same preference profile also have the same outcome in expectation. 
A randomized mechanism is \emph{envy-free} if it always selects an envy-free matching. 

An important property of a mechanism is \emph{strategy-proofness}. This means, that there is no incentive for any student not to submit her truthful preferences, no matter which preferences the other students report. 
A deterministic assignment mechanism $\chi$ is \emph{strategy-proof} if for any $\succeq \in \mathcal{P}^{|S|}$ with $i \in S$ and $\succeq_{i}' \in \mathcal{P}$ we have $\chi_{i}(\succeq) \succeq_{i} \chi_{i}(\succeq_{i}', \succ_{S \backslash \{i\} })$. It has been shown that participants in strategy-proof mechanisms such as the Vickrey auction do not necessarily bid truthfully in practice. Therefore, there was a recent discussion about obvious strategy-proofness of extensive form games \citep{li2017obviously}. Intuitively, a mechanism is obviously strategy-proof iff the optimality of truth-telling can be deduced without contingent reasoning. \cite{pycia2016obvious} show that RSD is a unique mechanism that is obviously strategy-proof, efficient, and symmetric in mechanisms without transfers.

%\begin{definition}[OSP, \citep{li2017obviously}]
%A strategy $\sigma$ is obviously dominant if, for all other strategies $\sigma'$, at any earliest information set where $\sigma$ and $\sigma'$ diverge, the best possible outcome from $\sigma'$ is no better than the worst possible outcome from $\sigma$. A mechanism is obviously strategy-proof (OSP) if it has an equilibrium in obviously dominant strategies.
%\end{definition}

For randomized mechanisms we need to adapt the definitions. A random assignment mechanism is (strongly) $SD$-strategy-proof if for every preference profile $\succeq$, and for all $i \in S$ and $\succeq'_i$ we have $\psi(\succeq_{i},\succeq_{-i}) \sdprefeq{i} \psi(\succeq_{i'},\succeq_{-i})$. 
A random assignment rule $\psi$ is \emph{weakly} $SD$-strategy-proof if for every preference profile $\succeq$, there exists no $\succeq'$ \emph{for some} student $i \in S$ such that $\psi(\succeq_{i'},\succeq_{-i}) \sdprefeq{i} \psi(\succeq_{i},\succeq_{-i})$. That is, there may not be any student $i$, who strictly prefers $\psi(\succeq_{i'},\succeq_{-i})$ over the truthful outcome, but there may be students $i$ who neither prefer $\psi(\succeq_{i'},\succeq_{-i})$ nor $\psi(\succeq_{i},\succeq_{-i})$. This can happen as the $\sdprefeq{}$-relation is not complete. We will omit the prefix $SD$ for brevity in the following. 
Note that there are also weaker notions of strategy-proofness for randomized mechanisms developed in the field of probabilistic social choice that we do not consider in this article. These notions are based on different ways of how to compare lotteries. Interested readers are referred to \cite{brandt2017rolling}.%Participation is another design goal that has received attention recently. \cite{brandl2017random} show that RSD and PS strictly incentivize single agents to participate, and both mechanisms can also cannot be manipulated by groups of agents who abstain strategically. We will not discuss participation further in this context.

In section \ref{sec:seconddes} we introduce a number of additional design goals that often matter in the practice and that we analyze empirically.

%In other words, an ordinal mechanism is strategy-proof if for any agent, the allocation resulting from misreporting is weakly stochastically dominated by the allocation from truthful reporting, with respect to an agent’s true preference. A mechanism is weakly strategy-proof if for each agent, his or her allocation from truthful reporting is not stochastically dominated by the allocation produced by a misreport, with respect to the agent’s true preference. 

\subsection{Assignment Mechanisms}
\label{sec:mechanisms}

A lot is known about assignment problems with single-unit demand. 
Random Serial Dictatorship (RSD) selects a permutation of the agents uniformly at random and then sequentially allows agents to pick their favorite course among the remaining ones. 
\cite{gibbard1977manipulation} showed that random dictatorship is the only anonymous and symmetric (in the sense of equal treatment of equals), strongly $SD$-strategy-proof, and ex post efficient assignment rule when preferences are strict. 
\cite{pycia2016obvious} prove that RSD is a unique mechanism that is obviously strategy-proof, efficient, and symmetric in mechanisms without transfers. 
In line with this recent result, \cite{AshlagiG15} show that stable matching mechanisms are not obviously strategy-proof.

% To see this, suppose three agents have von Neumann-Morgenstern preferences for lotteries over objects: Agent 1 has utility $u_1(a)=1$, $u_1(b)=0.9$, $u_1(c)=0$, and agent 2 has $u_2(a)=1$, $u_2(b)=0.2$, $u_2(c)=0$. Agent 3 is symmetric to agent 2. RSD gives a 1/3 probability of every object to each agent and expected utilities $(0.63, 0.4, 0.4)$. However, assigning object $b$ to agent 1 for sure and objects $a$ and $c$ randomly between agents $2, 3$ yields the expected utilities $(0.9, 0.5, 0.5)$. %%%%

However, RSD is not always ordinally efficient, only ex post efficient \citep{Bogomolnaia01}. \cite{zhou1990conjecture} actually showed that no random mechanism for assigning objects to agents can satisfy strong notions of strategy-proofness, ordinal efficiency, and symmetry simultaneously with more than three objects and agents. So, we also cannot hope for these properties in combinatorial assignment problems. RSD can also be applied to the combinatorial assignment problem. The Bundled Random Serial Dictatorship (BRSD) orders the students randomly and assigns the most preferred bundle which is still available to each student in this order. Although the package preferences take some toll on the runtime it is still very fast.

First-come first-served (FCFS) can be seen as a serial dictatorship. Students login at a certain registration and then reserve the most preferred bundle of courses that is still available. Although the arrival process is not uniform at random, students have little control over who arrives first. While there is a certain time when the registration starts, hundreds of students log in simultaneously to get course seats and it is almost random who arrives first. We will simulate FCFS via BRSD and run the algorithm repeatedly to get estimates for performance metrics of FCFS.

Probabilistic Serial (PS) \citep{Bogomolnaia01} produces an envy-free assignment with respect to the reported unit-demand preferences, and it is ordinally efficient, but it is only weakly $SD$-strategy-proof. 
Bundled Probabilistic Serial (BPS) by \cite{nguyen2015assignment} is a generalization of PS to the combinatorial assignment problem. BPS computes a fractional solution via a generalization of the PS mechanism. The BPS mechanism is also ordinally efficient, envy-free, and weakly strategy-proof if preferences are strict, which we will discuss as an issue in the conclusions. 

Informally, in BPS all agents \emph{eat} their most preferred bundle in the time interval $[0,1]$ simultaneously with the same speed as long as all included objects are available. As soon as one object is exhausted, every bundle containing this object is deleted and the agents continue eating the next available bundle in their preference list. The duration with which every bundle was eaten by an agent specifies the probability for assigning this bundle to this agent. 

\begin{algorithm}[htbp]
\DontPrintSemicolon
\SetKwFor{ForEach}{forall}{do}{}
\textbf{Input:} Preferences $(\succeq_{i})_{i\in S}$\;
$t = 0$\;
$x_{ib} =0$, $\forall i\in S, b\in B$\;
\While{$t<1$}{
	$D=\emptyset$\;
	$dem_j= 0$, $\forall j\in C$\;
	\lForEach{$i\in S$}{choose first valid bundle $b \in \succeq_{i}$: $D\leftarrow b$}
	\ForEach{$b\in D$}{
		\lForEach{$j\in b$}{$dem_j$++}
	}
	$\Delta = \mins{\frac{dem_j}{q_j}}{j\in C}$\;
	$t+=\Delta$\;
	$\Delta^\ast = \Delta-(t-\min{1,t})$\;	
	\lForEach{$i\in S$}{$x_{ib}+=\Delta^\ast$}
	\ForEach{$j\in C$}{
		$q_j -= \Delta^\ast\cdot dem_j$\;
		\lIf{$q_j = 0$}{$\forall b\in B: j\in b:$ delete $b$}
	}
	
}
\textbf{Output:} Allocation $x^\ast=(x_{ib})_{i\in S,b\in B}$
\caption{Pseudocode of BPS.}
\label{algo:bps}
\end{algorithm}
%TODO pseudocode beschreiben???

%Unfortunately, in contrast to the result of PS, the outcome of BPS is not implementable into a lottery of deterministic matchings in general. However, \cite{nguyen2015assignment} present a mechanism to decompose the BPS solution into a lottery over deterministic matchings, which over-allocate each course by at most $\ell-1$ seats, i.e. the \ref{eq:demand} constraints are fulfilled and only the \ref{eq:supply} constraints are relaxed. Details of the convex decomposition into integer solutions are provided in Appendix \ref{app:lottery}. 

%After a fractional solution was found via BPS, the optimal fractional solution $x^*$ of BPS is implemented as a lottery over integral matchings satisfying the \ref{eq:demand} constraints. We need to relax the right-hand side of the \ref{eq:supply} constraint by $l-1$ to be able to always implement the fractional solution as a lottery. Details of the convex decomposition into integer solutions are provided in \ref{sec:lottery}. 

\subsection{Implementing Random Assignments}
\label{sec:lottery}
Unfortunately, in contrast to the result of PS, the outcome of BPS is not implementable into a lottery of deterministic matchings in general if $\ell > 1$. To circumvent this, one can either scale $x^\ast$ by a factor $\alpha \in [0,1]$ such that the decomposition becomes possible \citep{Lavi11}
or one allows for the relaxation of some constraints. \cite{nguyen2015assignment} present a mechanism to decompose the BPS solution into a lottery over deterministic matchings, which over-allocate each course by at most $\ell-1$ seats, i.e. the \ref{eq:demand} constraints are fulfilled and only the \ref{eq:supply} constraints are relaxed. 

In the polynomial time \textit{lottery algorithm} (see Algorithm \ref{algo:lottery}), we find at most $d+1$ integral points, the convex hull of which is arbitrarily close to the fractional solution $x^\ast$, which we get from BPS. The lottery algorithm then returns a lottery over these $d+1$ integral vectors, which is close to $x^\ast$ in expectation. Variable $d$ describes the dimensions of the problem. In this lottery algorithm, we use a subroutine to return an integer point $\bar{x}$ such that $u\trans\bar{x} \geq u\trans x^*$. This subroutine is called \textit{iterative rounding algorithm}\index{iterative rounding} (IRA) and proceeds as described in Algorithm \ref{algo:ira}.

\begin{algorithm}[htbp]
\DontPrintSemicolon
\SetKwFor{Item}{}{}{}
\lItem{1a: }{Delete all $x_i = 0$, $x_i = 1$, update the constraints and go to 1b.}
%\SetKwBlock{Item}{1a:}{}
\Item{1b: }{If there is no $x_i\in \{1,0\}$ one can find at least one supply-constraint with
	$$\sum_{i \in \mathcal{S}}\sum_{b:j \in b}b_{j}\lceil x_{ib}\rceil \leq q_{j} + \ell - 1$$
	Delete those constraints and go to 2.
   }
%\SetKwBlock{Item}{2:}{}
\Item{2}{Solve $max\{u\trans x\;\mid\;(demand),(supply),x\in \mathbb{R}_{\geq 0}\}$\;
	\lIf{$\text{all }x_i \in \{0,1\}$}{return $x$}
	\lElse{go to 1a}
  }
\caption{Pseudocode of the iterative rounding algorithm.}
\label{algo:ira}
\end{algorithm}

%The constraint violation in step 2b. of the IRA is depicted in Figure \ref{fig:IRA}. 

%\begin{figure}
%	\centering
%		\includegraphics{../../figures/IRA.pdf}
%	\caption{Constraint violation in the IRA}
%	\label{fig:IRA}
%\end{figure}

Now, we can discuss the lottery algorithm (see Algorithm \ref{algo:lottery}). Let $\mathcal{B}(x^\ast, \delta)=\{x\;|\;\left |x^\ast -x\right |\leq \delta\}\subseteq \{x\in \mathbb{R}_{\geq 0} \;|\; (Demand)\}$ with $\delta > 0$. The parameter $\delta$ in $\mathcal{B}(x^\ast, \delta)$ determines some space around $x^\ast$ such that the demand constraint $\sum_{b\in B} x_{ib} \leq 1$ is not violated. It is always possible to determine such a $\delta$. If there is no slack in the demand constraints one has to scale down the fractional solution $x^\ast$. Afterwards one has to adjust the allowed error $\varepsilon$ such that after scaling and decomposition the original $\varepsilon$ is still fulfilled. Here, $\left |x - y\right |$ describes the Euclidean distance between two vectors $x$ and $y$. 

\begin{algorithm}[htbp]
\DontPrintSemicolon
\SetKwFor{Item}{}{}{}
\textbf{Input:} Fractional solution $x^\ast$\;
\lItem{1: }{Set $Z=\{\text{IRA}(x^\ast)\}$, i.e., find integer solution via IRA.} 
\Item{2: }{$y = argmin\{|x^\ast -y|\;|\; y \in conv(Z)\}$\; 
			\lIf{$|x^\ast -y|<\varepsilon$}{END}}
\lItem{3: }{Choose $Z' \subset Z$ of size $|Z'|\leq d$ and $y \in conv(Z'):\;$
			$z = x^\ast + \delta\frac{x^\ast -y}{|x^\ast -y|}$}
\lItem{4: }{Find optimal integral $z'$ s.t. (Demand),(Supply) and $(x^\ast -y)\trans z' \geq (x^\ast -y)\trans z$ via IRA}
\lItem{5: }{$Z = Z' \cup \{z'\}$ and go back to 2.}
\textbf{Output:} Convex combination of final $y$
\caption{Pseudocode of the lottery algorithm.}
\label{algo:lottery}
\end{algorithm}

%Step 3. of the lottery algorithm determines a new point $z$ by moving a $\delta$ in direction $x^\ast -y$. Starting from $z$ IRA is used to determine a close integral solution $z'$, which is then added to $Z$. The new hyperplane $(x^\ast -y)\trans z' \geq (x^\ast -y)\trans z$ is introduced when searching for $z'$. 

\begin{figure}[htb]
\centering
\includegraphics[scale=1]{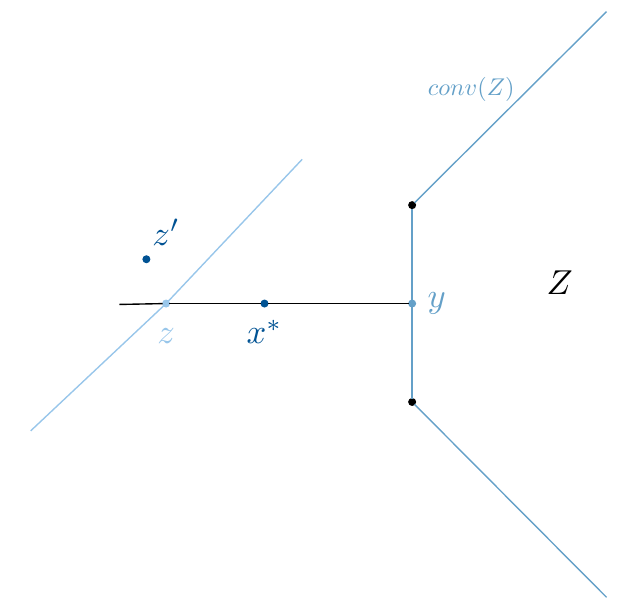}
\caption{Graphical representation of one iteration of the lottery algorithm.}
\label{fig:lottery}
\end{figure}

Figure \ref{fig:lottery} shows a graphical representation of one algorithm iteration. In each iteration the algorithm decreases the distance between $y$ and $x^\ast$ by adding a new integral solution to the solution set $Z$ and terminates when the distance between $y$ and $x^\ast$ is smaller than $\varepsilon$. That is, we consider $y$ as a good approximation for $x^\ast$ and return the support of $y$. The algorithm tries to get $x^\ast$ covered by the convex hull of $Z$ ($conv(Z)$). All solutions in $Z$ that are not part of the support of $y$, calculated in the quadratic optimization problem (QOP) in step $2$, are deleted (step 3). Thus, although we add a new integral solution to $Z$ in each iteration, the size of $Z$ never grows above $d+1$, since as long as $y\neq x^\ast$, $y$ always has to be on a face of $conv(Z)$. Hence, the support of $y$ consists of at most $d$ solutions. Step $4$ ensures that we search in the right direction for new integral solutions. As a side product the QOP also calculates the coefficients $\lambda^{(k)}$ for the convex combination and we have $x^\ast \approx y = \sum_{k=1}^{|Z|}\lambda^{(k)}x^{(k)}$, for $x^{(k)} \in Z$.

\section{Preference Elicitation}
\label{sec:matching}

This section focuses on the preference elicitation, which is important given the exponential set of possible bundles students might be interested in. We first introduce the environment and the problem for students, before we discuss different approaches to elicit their preferences.

\subsection{Background on the Application}

The Department of Informatics has been using stable matching mechanisms for the assignment of students to courses since 2014 \citep{diebold2017matching, diebold2014course}. The system provides a web-based user interface and every semester almost 1500 students are being matched to lab courses or seminars via the deferred acceptance algorithm for two-sided matching or random serial dictatorship for one-sided matching problems. 

In the context of the study reported in this paper, the web-based software was extended with BPS, the lottery mechanism for decomposing fractional solutions, and BRSD. 
$1439$ Students in computer science and information systems in their second semester participated in the matching during the summer term 2017 and they could choose tutorial groups from several courses including linear algebra, algorithms, software engineering, and operations research. A computer science student could have preferences for up to $5760$ $(=10\cdot 24\cdot 24)$ bundles\footnote{Consisting of the courses: linear algebra, algorithms, software engineering.} and an information systems student could have preferences for up to $5184$ $(=9\cdot 24\cdot 24)$ bundles.\footnote{Consisting of the courses: operations research, algorithms, software engineering.}
During the winter term 2017/2018, $1778$ computer science and information systems students in their third semester participated in the matching and could choose bundles of tutor groups out of four classes. A computer science student could have more than $700{,}000$ different bundles.\footnote{The computer science students need tutorials from all four classes $(< 22\cdot 25 \cdot 26 \cdot 52)$.}

\subsection{Automated Ranking of Packages}
\label{sec:auto_ranking}

A naive approach would be to let the students rank bundles on their own by choosing the time slots they want to have in their preference list. This would take a lot of time and lead to a substantial missing bids problem. 
We developed an algorithm that allows to rank-order all possible packages based on a few parameters that students need to specify. For this, we can leverage prior knowledge about timely preferences of students for schedules of tutorials and lectures.

Students' preferences mainly concern their commute and the possibility to free large contiguous blocks of time (e.g., a day or a half-day) that they can plan for other activites (e.g., a part-time job). 
In larger cities such as Munich, the time that students spend for commuting is significant. Also long waiting times between courses are perceived as a waste of time as it is often hard for them to work productively in several one- or two-hour breaks without appropriate office facilities available. 
For example, if a student had a tutorial on linear algebra in the morning, a lunch break, and then the tutorials for algorithms and software engineering in the afternoon of the same day with the minimal time for breaks specified, this would be considered ideal. The desired length for breaks between tutorials and for the lunch break are considered parameters with default values in the preference elicitation. 

\begin{figure}[!htbp]    
    \centering
    \includegraphics[width=0.75\textwidth]{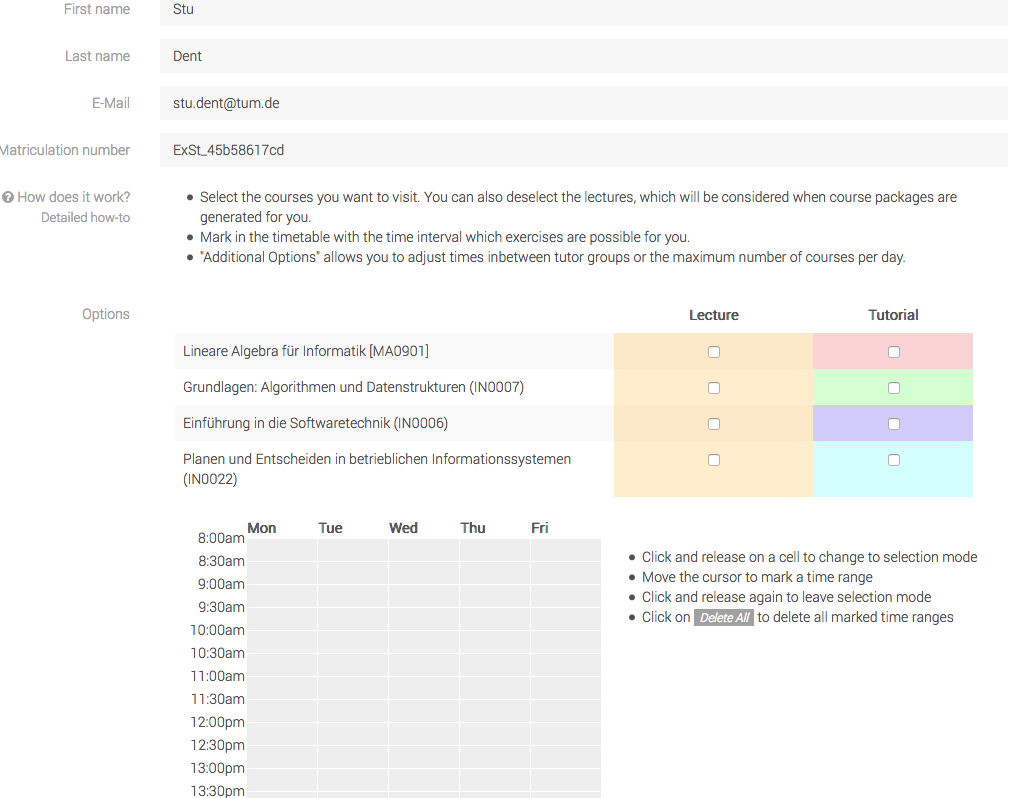}
    \caption{User Interface to Select Courses}
    \label{fig:prefoptions1}
\end{figure}

\begin{figure}[!htbp]
	\centering
    \includegraphics[width=0.60\textwidth]{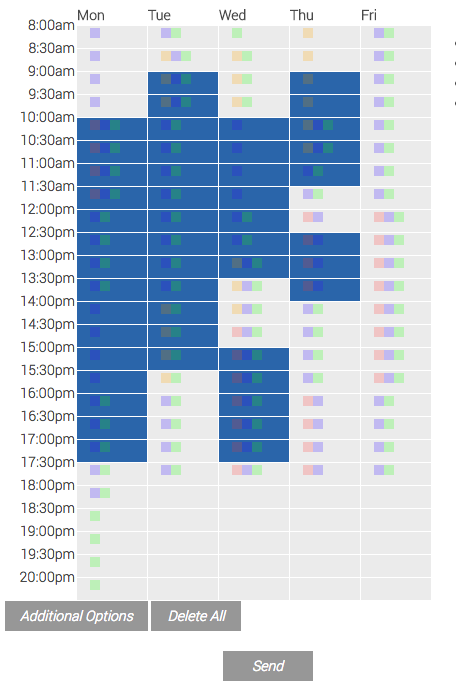}
    \caption{The Week Schedule}
    \label{fig:week_schedule}
\end{figure}

Figure \ref{fig:prefoptions1} shows the initial page where a student can select the courses of interest. On this page students choose the lectures and tutorials they are interested in. 
The selected lectures will be considered in the bundle generation as constraints, i.e. if a time slot of a tutorial overlaps with the time of a selected lecture, then it will no longer be considered in order to allow students to participate in the lecture. 
In a second step, the student marks available time ranges in a \emph{weekly schedule} (see Figure \ref{fig:week_schedule}). The day is partitioned into weekdays and time blocks of 30 minutes from 8:00 AM to 8:30 PM. If a tutorial is selected, all time slots of this tutorial will be highlighted with a specific color.
Thus, students learn when the tutorials and lectures of interest take place.

A student can set a minimal amount of time for a lunch break and a minimal amount of time in-between two events (default value is 15 minutes). We also allow students to provide weights $\{1,\ldots,5\}$ for the different days. That is, the students can express preferences over the days.

%Figure \ref{fig:additional_options} shows the relevant options.
% \begin{figure}[!htbp]
% 	\centering
%    \includegraphics[width=0.75\textwidth]{../../figures/additional_options.png}
%    \caption{Screenshot of the Additional Options}
%    \label{fig:additional_options}
%\end{figure}

The preferences elicited on this screen are input for an algorithm that uses prior knowledge about student preferences to rank-order all possible packages.   
The algorithm first generates bundles that satisfy all constraints and then ranks them. 
Finding the bundles that do not violate constraints (e.g., lectures to be attended) of the students can be cast as a \emph{constraint satisfaction problem}.  
After the feasible bundles are generated, we rank these bundles. For this we assign a score to each bundle that considers
\begin{itemize}
	\item how many days a student needs to come to the university per week in total,
	\item the preference ordering over the days,
	\item the total time a student has to stay at the university each day, and	
	\item the length of the lunch breaks between courses.% relative to the minimum break specified
\end{itemize}

The score for a package $b$ of courses across the week is the sum of the daily score $(score(b,d))$ for all weekdays $d$. The daily score is computed as

\begin{equation}
score(b,day) = \left (\dfrac{w(b,day)}{sp(b,day)}\cdot f(sp(b,day)) + br(b,day)\right )\cdot prio(day)
\end{equation}

This score is scaled between 0 and 27.5 at a maximum and it considers how well the day is utilized with courses. Ideally, a student would have all his tutorials on a single day, his most preferred day, have a 1-hour lunch break and a minimal time for breaks inbetween courses. This would yield 27.5 points.  

The time spent at the university per day $sp(b,day)$ is considered relative to the time a student attends courses on that day ($w(b,day)$). These courses include tutorials and lectures. The ratio is used to weigh the score for a day ($f(sp(b,day))$). This way a day where students do not spend more time in breaks than the minimum number of minutes specified maximizes the score. The function $f(\cdot)$ assigns 1 point for up to 2 hours spent at the university on a day ($sp(b,day) \leq 2$), 2 points for up to 4 hours, 6 points for up to 3 hours, 4 points for up to 8 hours, but only 2 points for days where a student is between 8 and 10 hours at the university. Longer schedules are not permitted.

A second component in the daily score $(score(b,d))$ is the lunch break. A 1-hour break was considered best. The scoring function $br(\cdot)$ would assign 0 points for lunch breaks less than 30 minutes, 1 point for 30-45 minutes, 1.5 points for 45-60 minutes, 2 points for 60-75 minutes, and again a low number of points for longer breaks. Students could also exclude schedules with a break less than a certain time, say 30 minutes. 

If the student does not have to visit the university at day $d$, he gets a fixed score of $30$ for day $d$. The daily scores are then multiplied by the priority of the day $[1..5]$. The overall score of a bundle $b$ is the sum of the $score(b,d)$ for all weekdays. 
As a result of this scoring rule, the more days the student can stay at home, the higher is the score of this bundle. As a simplified example, if a student had to come to the university on three different days to attend one course only, this bundle would get a score of less than 25, while if he could attend all courses on a single day with minimal breaks, this will get an overall score of more than 80 (for these three days).

In other words, the scoring rule will place bundles, that use a minimal number of days (ideally the most preferred days) with only a few breaks but a one hour lunch break on top of the preference list.
This would minimize the commute time and maximize the contiguous time a student can devote to learning or work. If the breaks between courses grow larger or courses take place on different or more days, this decreases the score. Ties are not impossible but almost never occur such that the algorithm typically generates a strict ranking of the feasible packages.

\hide{
\begin{table}
\centering
\begin{tabular}{|c|c|c|c|c|}\hline
\# courses & $w(b,day)$ (h) & lunch time (h)& $sp(b,day)(h)$ & $score(b,day)$\\ \hline
1 		   & 2			& 0			 & 2		   & $2\cdot prio(day)$\\
2 		   & 4			& 0			 & 4		   & $3\cdot prio(day)$\\
3 		   & 6			& 0			 & 6		   & $4\cdot prio(day)$\\
3 		   & 6			& 1			 & 7		   & $5.43\cdot prio(day)$\\
4 		   & 8			& 1			 & 9		   & $4.2\cdot prio(day)$\\\hline
\end{tabular}
\caption{assumption: 1 course = 2h}
\label{tab:scores}
\end{table}

} %%%%%%%% hide

\begin{figure}[!htbp]
	\centering
    \includegraphics[width=0.85\textwidth]{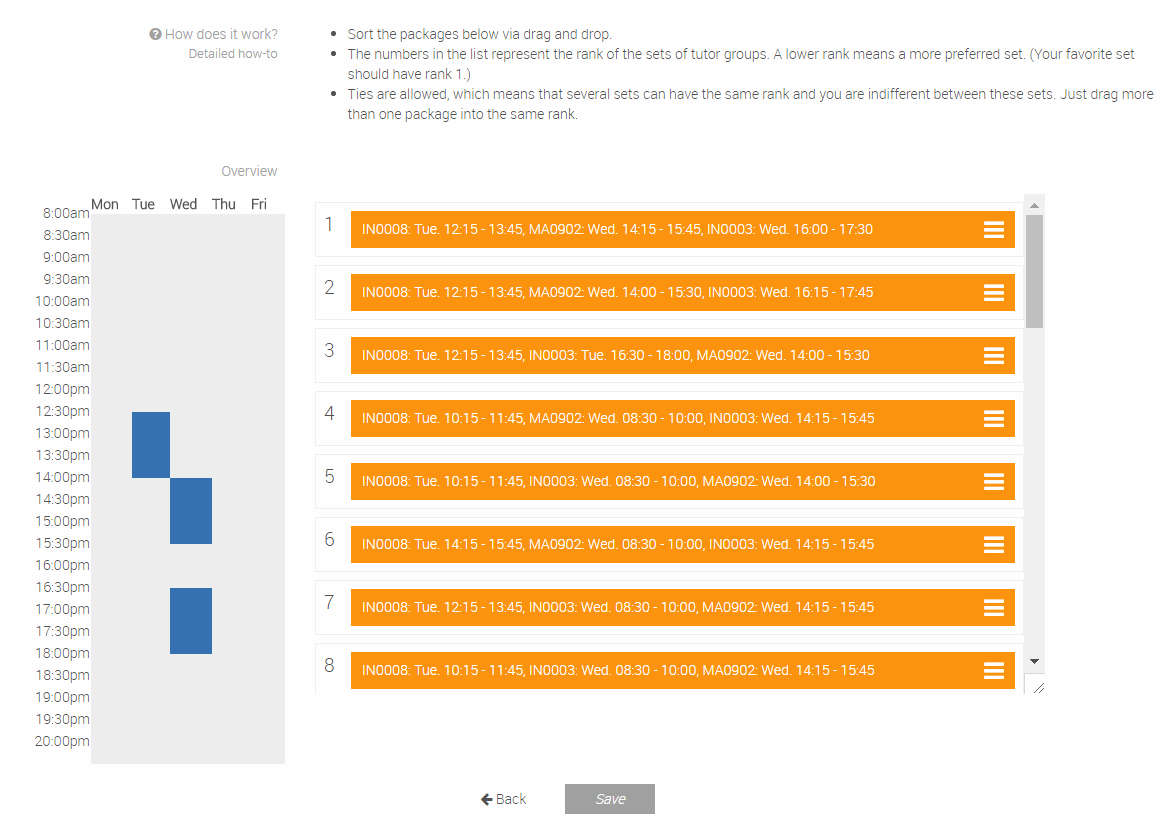}
    \caption{Page with top-ranked packages}
    \label{fig:ranking2}
\end{figure}

Even if it is a fair assumption that students have quite homogeneous preference structures, there might be some special cases we cannot cover with such a scoring rule. Therefore we give the students the possibility to adjust the outcome of this scoring procedure. On the ranking page, we display the $30$ top rated pre-ranked bundles and the students can adapt this ranking manually, go back to the previous screen and adapt the input parameters, or just accept the ranking with a single click (see Figure \ref{fig:ranking2}). Note that $\approx 90\%$ of the students received one of their top ten ranked packages and only a few students received a package with a rank less than $30$. %The probability of being matched to rank 30 in our experiments was less than $0.04\%$ and decreasing. 
So, if a student inspects and confirms the ranking of the first 10-30 packages, this covers the most important quantile of the overall ranking list. 
We generated a ranking over $200$ bundles for each student in advance based on the pre-specified parameters and further preferences only if necessary. 

\begin{figure}[!htbp]
	\centering
    \includegraphics[width=0.85\textwidth]{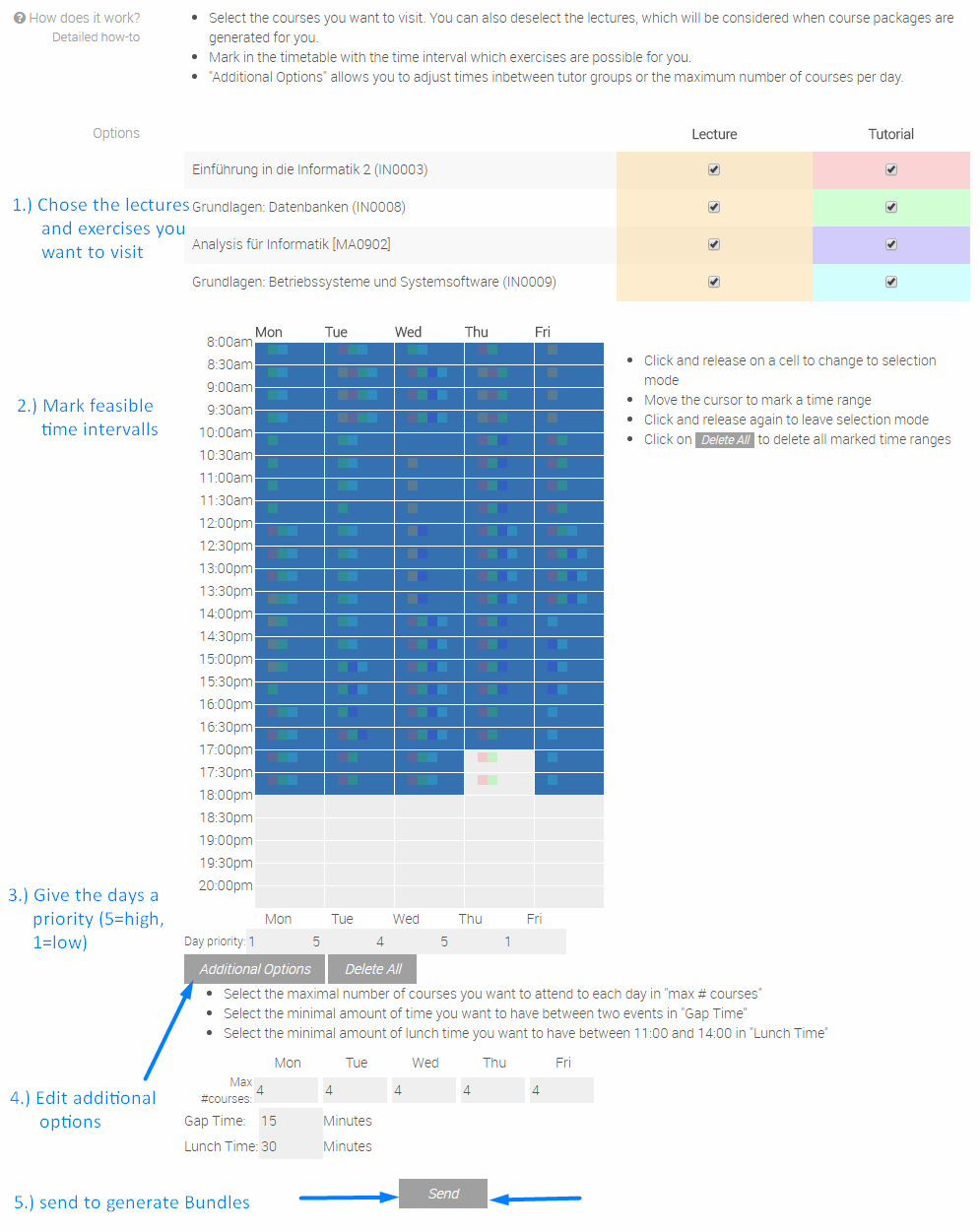}
    \caption{Process to rank-order packages}
    \label{fig:ranking3}
\end{figure}

So far, we described the user interface for the winter term 2017/18. 
The user interface in the summer term 2017 required students to explicitly drag and drop the pre-ranked packages on a screen. This was considered tedious such that in the winter term the generated ranking was suggested to students right away without any drag-and-drop activies required and could be confirmed without much effort. The main web page and the main steps a student had to take are summarized in Figure \ref{fig:ranking3}.

\subsection{Challenges of Course-Level Scoring}
\label{sec:pref_scoring}

Ranking an exponential set of packages is a general issue in course assignment problems, and one might ask if alternative methods are available. 
\cite{Budish2017} describes the preference elicitation used at the Wharton School of Business. Students could report cardinal item values on a scale of 1 to 100 for any course they were interested in taking. In addition, they could report adjustments for pairs of courses, which assigned an additional value to schedules that had both course sections together. Courses were then scored and transformed into an ordinal ranking over feasible schedules. The authors argue that they felt that ``adding more ways to express non-additive preferences would make the language too complicated.'' Wharton also provided a decision support tool listing the 10 most-preferred bundles, which allowed students to inspect top-ranked schedules and modify the cardinal values. 

Two problems make this method challenging to apply. 
First, the ranking is sensitive to minor changes in the weights, which is a well-known issue in multi-criteria decision making with additive value functions. Evaluation is characterized by a substantial degree of random error, and the amount of error tends to increase as the decision maker attempts to consider an increasing number of attributes (or courses in our case) \citep{bowman1963consistency, fischer1972four}. Difficulties in the calibration of scores for each course can lead to substantial differences in the resulting ranking.

Second, and more importantly, significant non-linearities arise due to the timely preferences of students in the assignment of tutorials, making it impossible to describe the preferences via a course-level utility function as proposed by \cite{Budish2017}. Even if three tutorials get a high number of points, this does not mean that their combination is preferable by a student as these tutorials might be on different days or have long breaks inbetween. To see this, we translated the ranking of packages into a set of inequalities with weights ($w$) as variables. Following revealed preference theory \citep{mas1995microeconomic}, we use these inequalities to understand whether there is any set of weights that would allow to describe the ranking using a utility function $\sum\limits_{i \in C} b_iw_i  +\sum\limits_{\substack{i,j \in C\\ j>i}} b_i b_j w_{ij}$. The function $r(b)$ describes the rank of a bundle, while $b$ is a binary parameter vector with each component $b_i \in \{0,1\}$ showing whether a course $i \in C$ is part of a package or not. The objective function minimizes the sum of error variables $\varepsilon$ in REV. If there is any set of weights that could reflect the ranking of packages in our experiments without these error variables, the resulting optimal objective value would be zero. For every violation of a constraint one has to increase the respective error variable to a positive value. 
%The resulting feasiblity problem should describe if there is any set of weights that could reflect the ranking of packages in our experiments. 

\begin{equation*} 
\begin{array}{rrcll} 
   \Min & err(\varepsilon) = \sum\limits_{b \in B} \varepsilon_b + \sum\limits_{\substack{i,j\in C\\ j>i}} \varepsilon_{ij}&&& \text{(REV)}\\
		& \sum\limits_{i \in C} b_iw_i  + \sum\limits_{\substack{i,j \in C\\ j>i}} b_i b_j w_{ij} 						+\varepsilon_{b}	&\geq& 
		\sum\limits_{i \in C} b'_iw_i  + \sum\limits_{\substack{i,j \in C\\ j>i}} b'_i b'_jw_{ij}																& \forall b,b': r(b')=r(b)+1\\
		& w_i + w_j + w_{ij} +\varepsilon_{ij} 	&\geq& 0 		& \forall i,j \in C, j > i \\
  		& w_{i} 				&\in &[0,1] 		&\forall i \in C\\
  		& w_{ij} 				&\geq& -2 		& \forall i, j \in C, j > i\\
  		& \varepsilon_{b},\varepsilon_{ij} 	 	&\geq& 0 & \forall i, j \in C, j > i, b\in B
\end{array}  
\end{equation*}

We solved the linear program for a large number of student preferences and in our environment, and none of the problems was feasible. We report results for a sample of students in Appendix \ref{app:sample_REV}. We had preferences ranking $4000$ to $12000$ bundles for the courses of the winter term.  None of these settings could be solved with objective value zero, that is, the generated preference lists are not representable with a linear model with adjustment-terms used by \cite{Budish2017}. Even if it was possible to find such a vector of course-level weights, it would probably be very difficult to parametrize by students. The way \cite{Budish2017} elicit preferences might be sufficient for settings, where students only are interested in a very small subset of groups of the courses. However, assuming that students are able to adjust weights for up to $50$ groups per course is utopian.

Eliciting preferences for hundreds of packages is a challenging problem, but the quality of any mechanism for CAP depends crucially on this input. There will be differences in the type of decision support one can provide in various types of applications. However, it is typically important that the timely preferences for students are captured.

\section{Results}

In Section \ref{sec:designdes} we have summarized first-order design goals for assignment problems: strategy-proofness, fairness, and efficiency. Now we introduce second-order design goals and respective metrics allowing us to compare the assignments of BPS and FCFS empirically. Then we provide numeric results and summarize the outcomes of a survey we conducted after the matching. 

\subsection{Metrics}
\label{sec:seconddes}

Apart from efficiency, fairness, and strategy-proofness, \emph{popularity} was raised as a design goal. An assignment is called popular if there is no other assignment that is preferred by a majority of the agents. 
Popular deterministic assignments might not always exist, but popular random assignments exist and can be computed in polynomial time \citep{kavitha2011popular}. 
However, \cite{brandt2017majority} prove that popularity is incompatible with very weak notions of strategy-proofness and envy-freeness, but it is interesting to understand the popularity of BPS vs. BRSD.
In our empirical evaluation we analyze whether BPS or FCFS are more popular. To measure popularity we first define the function $\phi_i(b,b'): B\times B \rightarrow \{\pm 1, 0\}$ associated with the preference relations:
\begin{equation}
\phi_i(b,b') = 
\begin{cases}
+1 & \text{if } b \succ_{i} b'\\
-1 & \text{if } b'\succ_{i} b\\
0  & \text{else}
\end{cases}
\end{equation}

\begin{definition}[Popularity]
\label{def:pop}
A random assignment $p\in \Delta$ is \emph{more popular} than an assignment $q$, denoted $p \blacktriangleright q$, if $pop(p,q) > 0$ with
\begin{equation}
pop(p,q) = \sum_{i\in S}\sum_{b,b'\in B}p_{ib}\cdot q_{ib'}\cdot \phi_i(b,b')
\end{equation}
A random assignment $p$ is \emph{popular}, if $\nexists q\in \Delta: q \blacktriangleright p$.
\end{definition}

Apart from popularity, the size and the average or median rank are of interest. The \emph{size} of a matching simply describes the number of matched agents. 
The \emph{average rank} is only meaningful in combination with the size of the matching, because a smaller matching could easily have a smaller average rank. We report the average rank, because it has been used as a metric to gauge the difference in welfare of matching algorithms in \cite{Budish2017} and \cite{Abdulkadiroglu-2009}, two of the few experimental papers on matching mechanisms. 

The \emph{profile} contains more information as it compares how many students were (fractionally) assigned to their first choice, how many to their second choice, and so on. The profile of two matchings is not straightforward to compare. 		
We want to compare multiple profiles based on a single metric, and decided to use a metric similar to the \emph{Area under the Curve of a Receiver Operating Characteristic} in signal processing \citep{hanley1982meaning} which was already used by \cite{diebold2017matching}. The \emph{Area Under the Profile Curve Ratio} (AUPCR) is the ratio of the Area Under the Profile Curve (AUPC) and the total area (TA) and is scaled between $0$ and $100\%$, where the AUPC describes the integral below the profile curve. 
The AUPCR up to a specific rank $n$ is equal to the probability that a matching mechanism will match a randomly chosen student higher than his $n$-th preference.

\begin{definition}[AUPCR \citep{diebold2017matching}]\label{def:AreaUnderProfileCurveRatio}
Let $C$ be the possible courses with $c\in C$ and $Q$ be the sum of all capacities, regarding the students $i\in S$ the AUPCR is defined as follows:
\begin{eqnarray*}
	TA\rbr{M} &=& \abs{C} \cdot \min{\abs{S}, Q} \\
	AUPC\rbr{M} &=& \sum_{r=1}^{\abs{C}} \abs{\{\rbr{i, c} \in M \mid \rank{i}{c} \leq r\}} \\
	AUPCR\rbr{M} &=& \frac{AUPC\rbr{M}}{TA\rbr{M}} \\
\end{eqnarray*}
\end{definition}

For the allocation of bundles we have to rewrite the definition of the AUPCR. 	
	
\begin{lemma}[AUPCR]\label{def:AreaUnderProfileCurveRatio}
    With $R$ denoting the number of possible ranks and $b\in B$, the AUPCR can be rewritten as:
   	\begin{eqnarray*}
		AUPCR\rbr{M} &=&\frac{1}{R} \sum_{r=1}^{R} \frac{\abs{\{\rbr{i,b} \in M \mid \rank{i}{b} \leq r\}}}{|S|}
	\end{eqnarray*}
\end{lemma}

Appendix \ref{sec:proof} provides a proof.
We have already introduced stochastic orders in Section \ref{sec:designdes}. We use second order stochastic dominance to compare two rankings \citep{levy1992stochastic}.%\footnote{In contrast to cumulative distribution functions, a more preferable profile is to the left and not to the right of a profile.}  

\subsection{Empirical Results}

The first application from the summer term 2017 comprised $1415$ students and $67$ courses (see Table \ref{tab:summer}). Overall, we had a list of $5847$ different bundles for the summer term. 
We simulated FCFS via BRSD on the preferences collected for the BPS. BPS is weakly strategy-proof and in such a large application it is fair to assume that students do not have sufficient information about the preferences of others. In the survey, we will see that a small proportion of the students reported that they deviated from truthful bidding and did not report some of their preferred time slots. However, taking the preferences for bundles of tutor groups elicited for the BPS allows for a comparison with BRSD. To compare the result of BPS and BRSD we actually would have to run the BRSD for all permutations of the students. Note that computing probabilities of alternatives in RSD explicitly is $\#P$-complete \citep{aziz2013computational}. We ran BRSD $1000$ to $1{,}000{,}000$ times with the same preferences but random permutations of the order of students and derived estimates for the different metrics. Since these results are very close, one can assume, that 1Mio runs of BRSD generate a good approximation to the (real) induced random matching.
%These estimates are very close (see Table \ref{tab:summer} and Table \ref{tab:winter}). 

\subsubsection{Popularity}

For the data from the summer and the winter term, BPS is more \emph{popular} than BRSD(1000000). $636$ students prefer BPS to FCFS, while $96$ students prefer FCFS to BPS. $683$ students are indifferent (see Table \ref{tab:popularity}). A positive popularity score as described in Definition \ref{def:pop} means, that BPS is more popular than the BRSD outcome and the score for BPS is $2.74$ for the summer term and $3.41$ for the winter term (compared to BRSD(1000000)). 
For the data from the winter term $754$ students prefer BPS to FCFS, while $120$ students prefer FCFS to BPS. $862$ students are indifferent (see Table \ref{tab:popularity}). Table \ref{tab:popularity} summarizes popularity and stochastic dominance for the summer and the winter term.  The syntax for the \emph{$SD$-preference} is the number of students preferring (BPS$|$BRSD(x)). It shows that BPS is preferable to BRSD according to $SD$-preference.

%%%%%%Data SS+WS
\begin{table}[htbp]
%\resizebox{\linewidth}{!}{
\centering
\begin{tabular}{|c|r|}
\hline Metric 		&  BRSD(1000000)\\ \hline 
popularity summer	&$2.73635$\\ 
popularity winter	&$3.41499$\\ 
$SD$-prefer summer	&$(636|96)$\\ 
$SD$-prefer winter	&$(754|120)$\\\hline
\end{tabular}%}
\caption{Popularity and stochastic dominance of BPS vs. BRSD}\label{tab:popularity}
\end{table}

\hide{
%%%%%%Data SS+WS
\begin{table}[htbp]
%\resizebox{\linewidth}{!}{
\centering
\begin{tabular}{|c|r|r|}
\hline Metric 		& BRSD(1000)& BRSD(1000000)\\ \hline 
popularity summer	&$2.35659$  &$2.73635$\\ 
popularity winter	&$1.93061$  &$3.41499$\\ 
$SD$-prefer summer	&$(611|273)$&$(636|96)$\\ 
$SD$-prefer winter	&$(690|299)$&$(754|120)$\\\hline
\end{tabular}%}
\caption{Popularity and stochastic dominance of BPS vs. BRSD}\label{tab:popularity}
\end{table}
}

\subsubsection{Rank and Size}

Table \ref{tab:summer} reports that in terms of average rank, average size, and the probability of being matched to one of the first 100 ranks BPS achieves higher scores in the summer term. Only the AUPCR for BRSD(1000000) is slightly better than for BPS.
The computation times were negligible for BRSD ($0.007$ seconds per run). BPS required $0.12$ seconds computation time with additional 6 minutes for the lottery algorithm in the summer term. This shows that BPS is a practical technique even for large assignment problems.

%%%%%%Data SS
%\begin{table}[htbp]
%%\resizebox{\linewidth}{!}{
%\centering
%\begin{tabular}{|c|r|r|r|}
%\hline Metric & BPS & BRSD(1000) & BRSD(1000000)\\ \hline 
%exp. rank 		&$2.20163$ &$2.20867$ &$2.20835$\\ 
%exp. size 		&$1086.58$ &$1085.84$ &$1085.79$\\ 
%prob. match (top 100)&$0.767901$ &$0.76738$ &$0.767345$\\ 
%AUPCR 			&$0.747419$ &$0.74679$ &$0.750782$\\ 
%weak envy 		&$0$ &$380$ &$381$\\ 
%strong envy	 	&$0$ &$981$ &$1064$\\ \hline
%\end{tabular}%}
%\caption{Summary statistics for the summer term 2017.}\label{tab:summer}
%\end{table}

%%%%%%Data SS (ohne BRSD1000)
\begin{table}[htbp]
%\resizebox{\linewidth}{!}{
\centering
\begin{tabular}{|c|r|r|}
\hline Metric & BPS& BRSD(1000000)\\ \hline 
exp. rank 		&$2.20163$ &$2.20835$\\ 
exp. size 		&$1086.58$ &$1085.79$\\ 
prob. match (top 100)&$0.767901$ &$0.767345$\\ 
AUPCR 			&$0.747419$ &$0.750782$\\ 
weak envy 		&$0$ &$381$\\ 
strong envy	 	&$0$  &$1064$\\ \hline
\end{tabular}%}
\caption{Summary statistics for the summer term 2017.}\label{tab:summer}
\end{table}

In the BPS outcome $72.735\%$ of the students receive an assignment ranked in their top ten while in BRSD $72.637\%$ receive such an outcome (see Table \ref{tab:profile_BPS_SS} for BPS and \ref{tab:profile_BRSD1000000_SS} for BRSD w. 1 mio. permutations of the students). Table \ref{tab:profile_BPS_SS} reports the probability of being matched to a particular rank and the AUPC in percentage for BPS, and Table \ref{tab:profile_BRSD1000000_SS} shows the rank profile for BRSD.

\begin{table}[htbp]\resizebox{\linewidth}{!}{
\begin{tabular}{|c|r|r|r|r|r|r|r|r|r|r|}\hline 
Rank  &$ 1$ &$ 2$ &$ 3$ &$ 4$ &$ 5$ &$ 6$ &$ 7$ &$ 8$ &$ 9$ &$ 10$\\ \hline% &$  \geq 11$\\ \hline
Prob match(\%) &$ 54.174$ &$  5.691$ &$  4.542$ &$  2.025$ &$  1.506$ &$  0.935$ &$  1.167$ &$  0.940$ &$  1.141$ &$  0.613$\\ \hline% &$  4.055$\\ \hline
%\# avr match &$ 766.56$ &$  80.53$ &$  64.27$ &$  28.66$ &$  21.31$ &$  13.23$ &$  16.52$ &$  13.31$ &$  16.14$ &$   8.67$ &$  57.38$\\ \hline
AUPC in (\%) &$ 54.174$ &$ 59.865$ &$ 64.407$ &$ 66.432$ &$ 67.938$ &$ 68.874$ &$ 70.041$ &$ 70.981$ &$ 72.122$ &$ 72.735$\\ \hline% &$ 76.790$\\ \hline
%AUPC in \# stud &$ 766.56$ &$ 847.09$ &$ 911.36$ &$ 940.02$ &$ 961.33$ &$ 974.56$ &$ 991.08$ &$ 1004.38$ &$ 1020.53$ &$ 1029.20$ &$ 1086.58$\\\hline
\end{tabular}}
\caption{Rank profiles for BPS in summer term 2017.}\label{tab:profile_BPS_SS}
\end{table}

\begin{table}
\resizebox{\linewidth}{!}{
\begin{tabular}{|c|r|r|r|r|r|r|r|r|r|r|}\hline 
Rank  &$ 1$ &$ 2$ &$ 3$ &$ 4$ &$ 5$ &$ 6$ &$ 7$ &$ 8$ &$ 9$ &$ 10$\\ \hline% &$  \geq 11$\\ \hline
Prob match(\%) &$ 53.973$ &$  5.725$ &$  4.538$ &$  2.053$ &$  1.529$ &$  0.931$ &$  1.181$ &$  0.948$ &$  1.150$ &$  0.610$ \\ \hline%&$  4.097$\\ \hline
%\# avr match &$ 763.71$ &$  81.01$ &$  64.22$ &$  29.05$ &$  21.64$ &$  13.17$ &$  16.71$ &$  13.41$ &$  16.27$ &$   8.64$ &$  57.98$\\ \hline
AUPC in (\%) &$ 53.973$ &$ 59.697$ &$ 64.236$ &$ 66.289$ &$ 67.818$ &$ 68.748$ &$ 69.929$ &$ 70.877$ &$ 72.027$ &$ 72.637$\\ \hline% &$ 76.734$\\ \hline
%AUPC in \# stud &$ 763.71$ &$ 844.72$ &$ 908.93$ &$ 937.98$ &$ 959.62$ &$ 972.79$ &$ 989.50$ &$ 1002.91$ &$ 1019.18$ &$ 1027.82$ &$ 1085.79$\\\hline
\end{tabular}
}\caption{Rank profile BRSD(1000000) in summer term 2017.}
\label{tab:profile_BRSD1000000_SS}%
\end{table}

%\subsubsection{Winter term}

The second application in the winter term included $1736$ students and $66$ courses. Overall, we had a list of $20{,}845$ different bundles for the winter term. Again, BPS achieved better results than BRSD in all metrics (see Table \ref{tab:winter}).  

In the BPS outcome $89.047\%$ of the students receive an assignment ranked in their top ten while in BRSD $88.891\%$ receive such an outcome (see Table \ref{tab:profile_BPS_WS} for BPS and \ref{tab:profile_BRSD1000000_WS} for BRSD with 1 mio. permultations of the students).
The computation times were again very low. BPS required $0.382$ seconds, but the lottery algorithm around $30$ minutes due to the higher number of bundles generated in the winter term.

%%%%%%Data WS
%\begin{table}[htbp]
%%\resizebox{\linewidth}{!}{
%\centering
%\begin{tabular}{|c|r|r|r|}
%\hline Metric & BPS & BRSD(1000) & BRSD(1000000)\\ \hline 
%exp rank &$1.97372$ &$1.9784$ &$1.97873$\\ 
%exp size &$1603.01$ &$1601.03$ &$1600.84$\\ 
%prob match (top 100)&$0.923394$ &$0.922253$ &$0.922142$\\ 
%AUPCR &$0.889512$ &$0.888184$ &$0.888058$\\ 
%weak envy &$0$ &$427$ &$451$\\ 
%strong envy &$0$ &$1050$ &$1202$\\ \hline
%\end{tabular}%}
%\caption{Summary statistics for the winter term 2017/2018.}\label{tab:winter}
%\end{table}

%%%%%%Data WS ohne BRSD1000
\begin{table}[htbp]
%\resizebox{\linewidth}{!}{
\centering
\begin{tabular}{|c|r|r|r|}
\hline Metric & BPS & BRSD(1000000)\\ \hline 
exp rank &$1.97372$&$1.97873$\\ 
exp size &$1603.01$ &$1600.84$\\ 
prob match (top 100)&$0.922253$ &$0.922142$\\ 
AUPCR &$0.889512$ &$0.888058$\\ 
weak envy &$0$ &$451$\\ 
strong envy &$0$ &$1202$\\ \hline
\end{tabular}%}
\caption{Summary statistics for the winter term 2017/2018.}\label{tab:winter}
\end{table}

\subsubsection{Envy}

Our experiments in the summer and the winter term confirm the theoretical result that BPS is (strongly) \emph{envy-free}. BRSD is neither weakly nor strongly envy-free. In the summer term, $1064$ students do not fulfill the envy-freeness condition (see Definition \ref{def:envy}), from which $381$ students do not even fulfill the weak envy-freeness condition (see BRSD(1000000) in Table \ref{tab:summer}). Similarly, for the winter term $1202$ students do not $SD$-prefer their outcome over the outcomes of every other student, and $451$ of those students even prefer an outcome of another student (see BRSD(1000000) in Table \ref{tab:winter}).

\begin{table}[htbp]\resizebox{\linewidth}{!}{
\begin{tabular}{|c|r|r|r|r|r|r|r|r|r|r|}\hline 
Rank  &$ 1$ &$ 2$ &$ 3$ &$ 4$ &$ 5$ &$ 6$ &$ 7$ &$ 8$ &$ 9$ &$ 10$\\ \hline% &$  \geq 11$\\ \hline
Prob match(\%) &$ 73.596$ &$  7.083$ &$  3.392$ &$  1.660$ &$  1.041$ &$  0.698$ &$  0.465$ &$  0.447$ &$  0.366$ &$  0.299$\\ \hline% &$  3.293$\\ \hline
%\# avr match &$ 1277.62$ &$ 122.96$ &$  58.89$ &$  28.81$ &$  18.08$ &$  12.12$ &$   8.07$ &$   7.76$ &$   6.35$ &$   5.20$ &$  57.16$\\ \hline
AUPC in (\%) &$ 73.596$ &$ 80.678$ &$ 84.070$ &$ 85.730$ &$ 86.772$ &$ 87.470$ &$ 87.935$ &$ 88.381$ &$ 88.747$ &$ 89.047$\\ \hline% &$ 92.339$\\ \hline
%AUPC in \# stud &$ 1277.62$ &$ 1400.58$ &$ 1459.46$ &$ 1488.28$ &$ 1506.36$ &$ 1518.48$ &$ 1526.55$ &$ 1534.30$ &$ 1540.65$ &$ 1545.85$ &$ 1603.01$\\\hline
\end{tabular}}
\caption{Rank profiles for BPS in winter term 2017/2018.}\label{tab:profile_BPS_WS}
\end{table}

\begin{table}
\resizebox{\linewidth}{!}{
\begin{tabular}{|c|r|r|r|r|r|r|r|r|r|r|}\hline 
Rank  &$ 1$ &$ 2$ &$ 3$ &$ 4$ &$ 5$ &$ 6$ &$ 7$ &$ 8$ &$ 9$ &$ 10$\\ \hline%  &$  \geq 11$\\ \hline
Prob match(\%) &$ 73.452$ &$  7.046$ &$  3.382$ &$  1.673$ &$  1.040$ &$  0.704$ &$  0.486$ &$  0.443$ &$  0.358$ &$  0.307$\\ \hline%  &$  3.323$\\ \hline
%\# avr match &$ 1275.12$ &$ 122.31$ &$  58.71$ &$  29.05$ &$  18.06$ &$  12.22$ &$   8.44$ &$   7.69$ &$   6.21$ &$   5.33$ &$  57.69$\\ \hline
AUPC in (\%) &$ 73.452$ &$ 80.497$ &$ 83.879$ &$ 85.553$ &$ 86.593$ &$ 87.297$ &$ 87.783$ &$ 88.226$ &$ 88.584$ &$ 88.891$\\ \hline%  &$ 92.214$\\ \hline
%AUPC in \# stud &$ 1275.12$ &$ 1397.44$ &$ 1456.14$ &$ 1485.19$ &$ 1503.25$ &$ 1515.47$ &$ 1523.92$ &$ 1531.61$ &$ 1537.81$ &$ 1543.15$ &$ 1600.84$\\\hline
\end{tabular}
}\caption{Rank profile BRSD(1000000) in winter term 2017/2018}
\label{tab:profile_BRSD1000000_WS}%
\end{table}

\subsection{The Lottery of the Summer Term Instance}
%TODO
We already have discussed, that we still have to decompose the solution of BPS into a lottery over integral solutions, to choose a deterministic allocation. This subsection presents exemplary with the Data from summer term 2017, how such a lottery is structured, and how significant the problem of overallocation is in practice. 

\begin{figure}[!htbp]
    \includegraphics[width=0.9\linewidth]{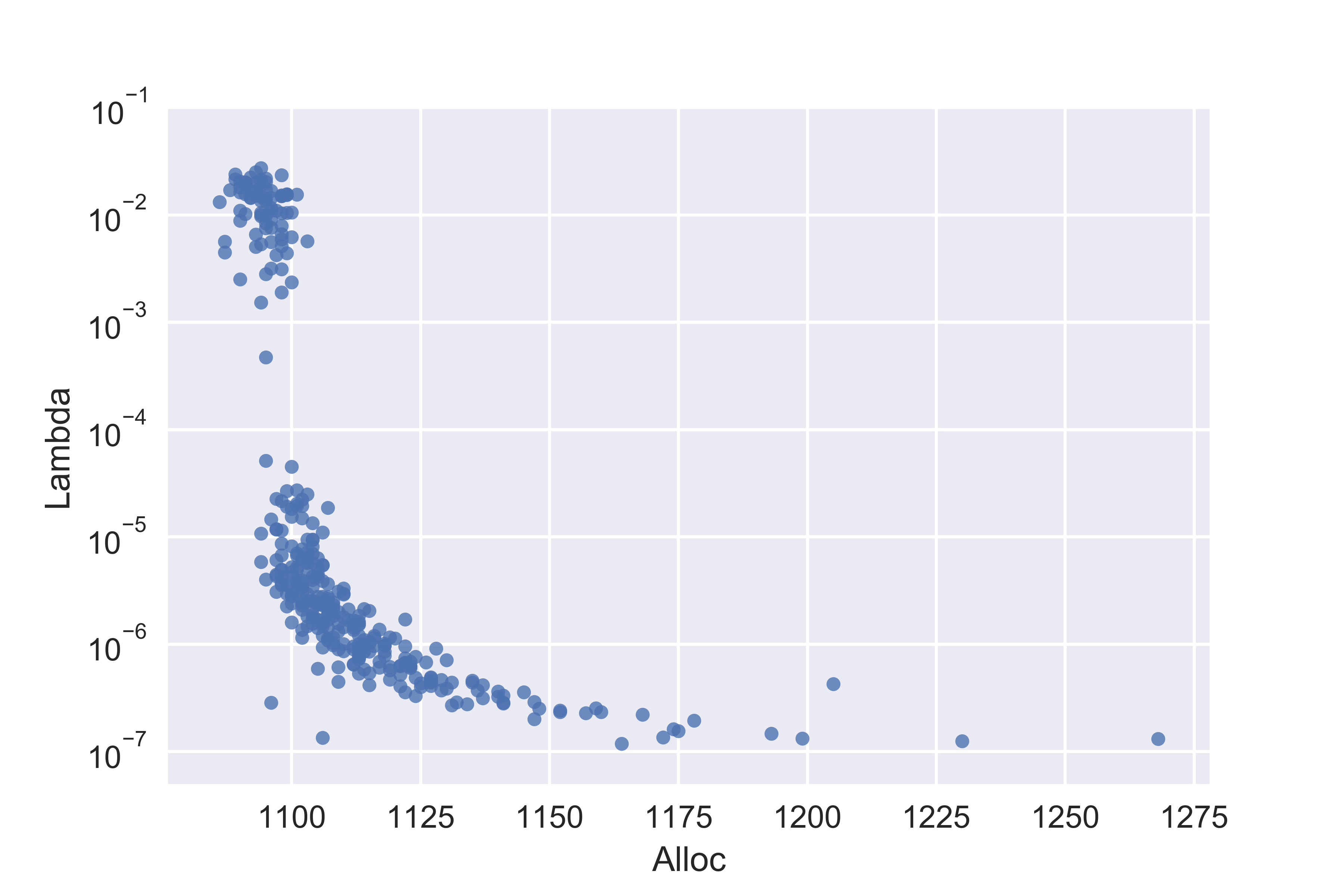}
    \caption{The lottery: Probabilities ($\lambda$) and size of the different deterministic matchings returned by Algorithm \ref{algo:lottery}.}
    \label{fig:alloclamb}
\end{figure}
Figure \ref{fig:alloclamb} shows the lottery resulting from decomposing the BPS solution described in \ref{tab:summer} into a lottery over approximative feasible integral solutions via Algorithm \ref{algo:lottery}.
We see that most solutions are close to the fractional solution in terms of number of allocated students (remember: the size of the BPS solution is $1086.58$).

One interesting question is how the solutions with a bigger size differ from the matchings with a lower number of allocated students. We computed the average ranks of the deterministic solutions and compared them with the size of the particular matchings.
\begin{figure}[!htbp]
    \includegraphics[width=0.9\linewidth]{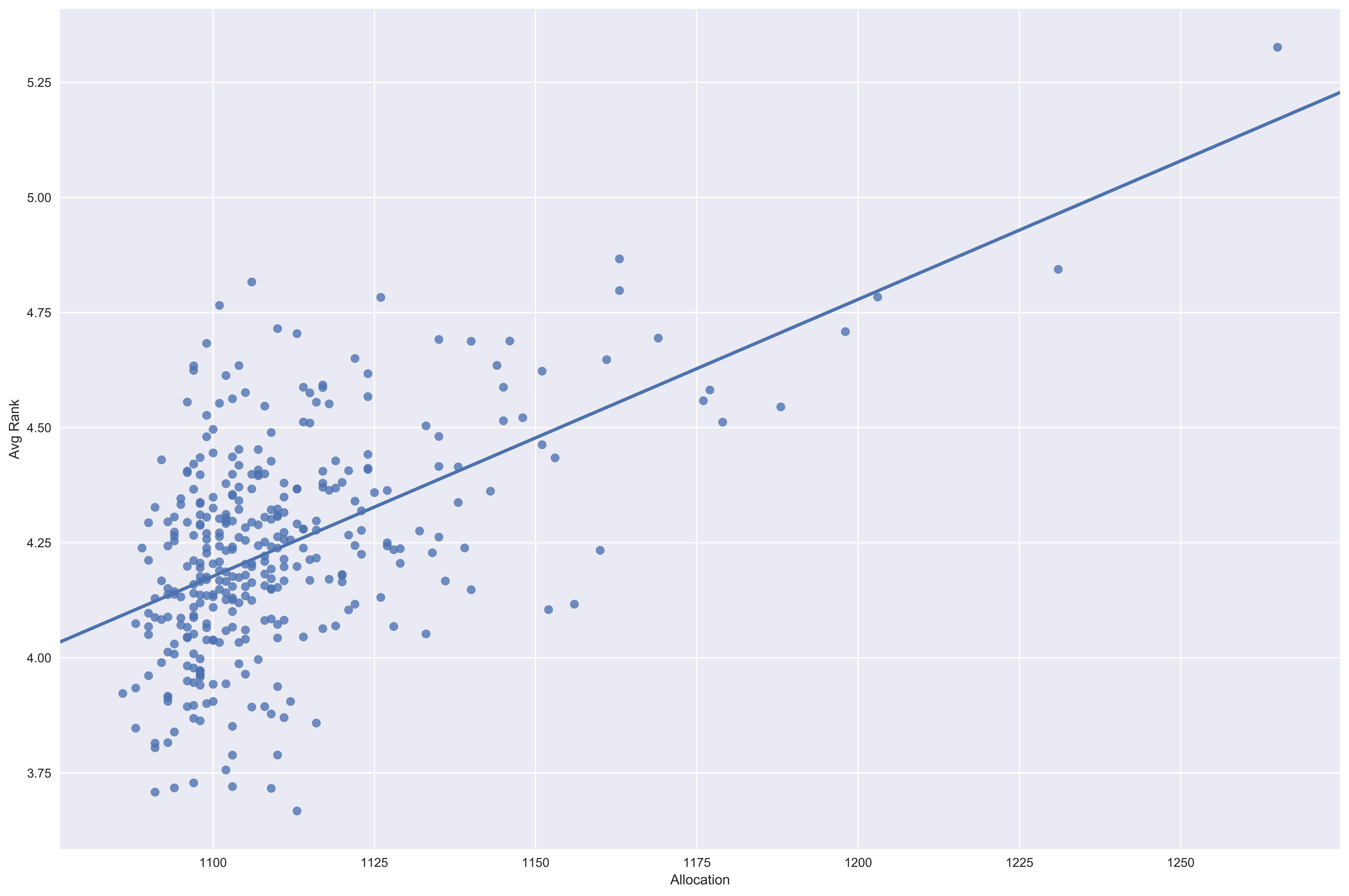}
    \caption{Average rank vs. size of the matchings in the lottery.}
    \label{fig:quali}
\end{figure}
Figure \ref{fig:quali} shows the distribution of the different matchings in the lottery with respect to size and  average rank. In the first fourth of the x-axis (the size of the matching) the variance is high for the average rank but the pattern gets clearer for a size higher than $1150$ -- there is a trade-off between size of the matching and the average rank of the allocated students.

In section \ref{sec:mechanisms} and \ref{sec:lottery} we already discussed that the capacity constraints of the courses might be violated. In the reported instance, $\ell = 4$. Hence, the worst case violation of these constraints is $3$. 

For the computation, we run the lottery with $\epsilon = 2.0$.
Let $e_{x^{(k)},L}$ be the number of goods that experienced a supply violation by $L$ units in the (integral) solution $x^{(k)}$, $\lambda^{(k)}$ the probability of matching $x^{(k)}$ and 
$$E_L(Z) = \sum_{x^{(k)}\in Z} \lambda^{(k)}\cdot e_{x^{(k)},L}$$
shows us how often an over allocation of exact $L$ seats occurs in the set of the lottery $Z$ on average.
With the settings mentioned above we receive:

\begin{itemize}
\item $E_1(X) = 5.36$
\item $E_2(X) = 0.64$
\item $E_3(X) = 0.07$
\end{itemize}

We see that an over allocation by $3$ seats rarely happens. Even a violation of $2$ seats occurs on average only in $0.64$ of the $67$ time courses. An over allocation of $1$ seat occurs on average in $5.36$ courses. In the actual application this overallocation did not even require special procedures and course organizers could typically accommodate one or two more students without problems.

The violations also barely change for results with $\epsilon = 1.0$:
\begin{itemize}
\item $E_1(X) = 5.23$
\item $E_2(X) = 0.9 $
\item $E_3(X) = 0.06$
\end{itemize}

We informed the course organizer before the matching, that small violations of the capacities are possible and no one had a problem with that. If the capacities of some courses were tight, on could solve the problem, by defining a smaller capacity for those courses. Theoretically one had to reduce the capacity by $\ell-1$ (3 in our case). However, our empirical results suggest, that a reduction of one seat should be sufficient, to ensure a feasible allocation with a high probability. The reduction of the capacities; however, come of cost of a lower efficiency of the matching in general.

\subsection{Survey Results}

After the students were assigned to the tutor groups and the courses started, we conducted a survey among the students using a 5-point Likert scale ($1 =$ strongly agree, $2 =$ agree, $5 =$ strongly disagree). 169 students out of 1736 students participated in the survey in the winter term and we report their responses in Table \ref{tab:survey}. Note that the students were exposed to FCFS in other semesters and now participated in BPS, which allowed them to compare both mechanisms. 

Students neither had to participate and we made clear that the feedback was used for research purposes only. 
The responses indicate that the majority of the students responding found the system easy to use and that they could express their preferences well. More than 50\% agreed (2) or strongly agreed (1) to questions 1 to 6. A majority also considers the system as fair (question 7), but almost 22\% of the respondents also disagreed to this statement. Note that students might have had an understanding of fairness that is different from envy-freeness or equal treatment of equals. For example, some students felt that in FCFS they could improve their assignment by making sure that they are among the first to register. This was perceived as fair as the additional effort would lead to higher chances of getting their best allocation as compared to those students who do not care about the assignment as much. 

%\begin{sidewaystable} 
\begin{table}
%\begin{center} 
\resizebox{\linewidth}{!}{  
\begin{tabular}{|r|l|r|r|r|r|r|}\hline
   &\textbf{Question}  & \textbf{1} & \textbf{2} & \textbf{3} & \textbf{4} & \textbf{5} \\ \hline 
  1& I had no problems to select my time ranges in the weekly schedule & $34.9$ & $34.9$ & $11.8$ & $9.5$ & $8.9$ \\
  2& The ranking of the generated sets of time slots was easy & $26.6$& $26.6$ & $18.9$ & $14.8$ & $13.0$ \\
  3& The instructions on the matching system were sufficient & $25.4$ & $37.3$ & $18.3$ & $10.1$ & $8.9$ \\
  4& The generated sets of tutorial groups met my expectations & $37.9$ & $27.8$ & $10.1$ & $9.5$ & $14.8$\\
  5& I was able to express my preferences on sets of tutor groups well & $42.6$ & $24.9$ & $13.6$ & $7.7$ & $11.2$ \\
  6& I consider the way bundles are allocated through the matching system as fair & $32.5$ & $27.2$ & $18.3$ & $5.9$ & $16.0$ \\
  7& I am satisfied with the matching outcome & $45.0$ & $17.0$ & $9.5$ & $6.5$ & $21.9$ \\
  8& I felt like I had control over my schedule & $29.0$ & $18.9$ & $13.0$ & $17.2$ & $21.9$ \\
  9& I was expressing my preferences truthfully & $72.4$ & $13.4$ & $4.2$ & $3.6$ & $6.5$ \\
  10& I was strategically hiding some of my most preferred time slots & $5.3$ & $4.7$ & $8.3$ & $13.6$ & $68.0$ \\
  11& I was strategically hiding some of my least preferred time slots & $16.0$ & $12.4$ & $16.0$ & $12.4$ & $43.2$ \\ \hline 
\end{tabular} 
}\caption{Survey results, values in \%}
\label{tab:survey}
%\end{center} 
\end{table}
%\end{sidewaystable}

62.1\% of the respondents were satisfied with the outcome (agreed or strongly agreed), while 28.4\% were not. It is unclear how those students who did not respond perceived the outcome, but there is a tendency that students, who are unhappy with the outcome, rather respond than students who got a high ranked bundle. Hence, the sample of students who respond might be slightly biased towards unsatisfaction. The ranking and profile information reported earlier provides alternative information about satisfaction of students with the outcome.

85.8\% of the respondents reported that they were expressing their preferences truthfully in BPS (agreed or strongly agreed), while around 10.1\% did not (disagreed or strongly disagreed). 10.1\% were also indicating that they were hiding some of their most preferred time slots, while even 28.4\% agreed or strongly agreed to the statement that they were hiding some of their least preferred time slots. This high percentage needs to be seen in conjunction with the exponentially large set of possible packages. If a student provides many possible time slots, then the list of packages grows very large. Therefore, there might have been a tendency to narrow down the selection of acceptable time slots, i.e. not rank the least preferred time slots. 

Still, the fact that a significant part of the students indicate that they did not report preferences truthfully is a tangible difference to FCFS. In FCFS, students only provide their single best package at the point in time, when they log in. This is simple, intuitive, and obviously strategy-proof. This property has to be traded off against the level of envy in BPS.

In a final question students were asked whether they prefer FCFS or BPS: 106 students (62,7\%) preferred BPS, while 63 (37.3\%) preferred FCFS. To understand the concerns of those students who preferred FCFS, it is useful to look at the written comments.  
Some students who provided comments were unhappy with the outcome, others were unhappy about the effort to rank-order their packages. 
%For example, one student suggested to eliminate the ranking of specific bundles and only provide feasible time slots. 

%Some of the students complained that not all of their courses could be considered in the matching. This means some students took additional classes that were not part of the assignment.  
% A request that was raised repeatedly was that students want to study in groups and they would like to have the same schedule. 

\subsection{Discussion of Differences}

The results from our field experiments and the survey reveal a number of interesting insights. Overall, BPS dominates BRSD on all metrics from our empirical evaluation in both field studies. It has a better average rank, a higher average size and a higher probability of matching, and it does not exhibit envy. 
However, the differences in average rank, average size, and the profile curve (AUPCR) are small, which is interesting given the fact that only a small number of preferences per student are considered via FCFS.  

There are a number of reasons that help explain the close performance of BPS and FCFS in these metrics. 
First, \cite{che2010asymptotic} find that random serial dictatorship and probabilistic serial become equivalent when the market becomes large, i.e. the random assignments in these mechanisms converge to each other as the number of copies of each object type grows, and the inefficiency of RSD becomes small. Our empirical results suggest that differences might also be small in large combinatorial assignment markets with limited complementarities. 

Second, ordinal preferences do not allow to express the intensity of preferences. Suppose there are two students who both prefer course $c_1$ to $c_2$, each having one course seat only. No matter who gets course $c_1$, the average rank and size of the matching as well as the profile will be the same even though one student might desperately want to attend $c_1$, while the second student only has a mild preference for $c_1$. Without cardinal information about the intensity of a preference the differences in aggregate metrics can be small. 

Third, an earlier comparison of FCFS with a deferred acceptance algorithm by \cite{diebold2014course} also showed that FCFS yielded surprisingly good results. While the average rank of FCFS was worse, the size of the matching resulting from FCFS was significantly larger compared to that from the deferred acceptance algorithm. For the combinatorial assignment problem, BPS actually had a larger average size than FCFS in both studies. For applications of matching in practice it is important to understand these trade-offs.

%For example, in applications where one agent could demand all objects envy would be very high. 
% A comment of caution is in order with respect to the random tie breaking. In BPS we have an exponentially long list of packages and most of these packages are not being ranked. For (B)PS it is necessary to randomly break ties. It might well be possible that efficiency and envy could be improved by considering ties.

%\section{Related Literature}
%Randomization has played an increasingly important role in market design. 
%RSD is strategy-proof and Pareto optimal but lacks ordinal or ex ante efficiency and envy-freeness. This is particularly troublesome if agents have multi-unit demands, where the first agent picked randomly might get all objects and others get none. Other random allocations might give greater expected utility.
%
%Therefore, there has been an interest in ex-ante efficient and equitable mechanisms, regardless of strategy-proofness. Examples are PS \citep{bogomolnaia2001new} and CEEI \citep{Hylland1979}. These mechanisms specify probability shares in objects and not lotteries over feasible outcomes. PS is ordinally efficient and envy-free, but only weakly strategy-proof. 

\section{Conclusions}

We report two large field studies and show that BPS performs well on a number of additional criteria including average rank, average size, probability of a matching among the first 100 ranks, and the overall profile of ranks (in terms of AUPC of a specific rank) assuming a complete, truthful, and strict ranking of all packages. The matching based on BPS is also more popular than BRSD based on the preferences submitted for BPS. The level of envy in FCFS is significant, even though the size of the packages that can be submitted is limited to the number of classes (three to four groups per package). 

The assignment of tutor groups is specific as preferences are mainly about times of the week. The preferred time slots in a week are different from student to student. However, the way how tutor groups should be ordered within these time slots (e.g., time for breaks) can be described with a few parameters such that it was possible to generate packages according to a score. The feedback of students was that this automated ranking met their preferences well and we argue that this is a good way to address the missing bids problem in similar applications. In other applications, generating good bundles might not be as straightforward and this will have an impact on efficiency. Compact and domain-specific bid languages have been discussed in the auction literature \citep{Bichler11a}, and they could also be a possibility to allow mechanisms without transfers circumvent the missing bids problem. 

The paper highlights basic trade-offs in market design without money: FCFS can be seen as a version of serial dictatorship which is ex post efficient, and obviously strategy-proof and treats students equally. It is also transparent and simple to implement and understand for students. BPS is a new randomized mechanism that is only weakly strategy-proof, but envy-free, and ordinally efficient, which is stronger than ex-post efficiency assuming strict preferences. Note that these properties hinge on the availability of strict preferences over all, exponentially many, bundles. 

Even if the missing bids problem can be addressed, two important problems remain: First, in contrast to FCFS the BPS mechanism is not obviously strategy-proof and a part of the students in the survey already indicated that they either hid their most preferred or least preferred time slots strategically.\footnote{Remember that our empirical comparisons are based on the preferences reported in BPS. A part of these preferences might not have reflected the true preferences of participants, and the comparison might be biased towards BPS.} Second, the assumption of strict preferences is strong in the presence of exponentially many bundles. Unfortunately, extending PS or BPS to preferences with ties is not without loss. On the one hand, \cite{katta2006solution} extended PS to preferences with indifferences and showed that it is not possible for any mechanism to find an envy-free, ordinally efficient assignment that satisfies even weak strategy-proofness as in the strict preference domain. On the other hand, with indifferences and random tie breaking efficiency cannot be guaranteed. Our preference elicitation technique generated a strict and complete ranking of course bundles based on a few input parameters and is one way to address these issues.

The key difference between BPS and FCFS is the absence of envy. The level of envy in FCFS is significant. Note that it might be even more pronounced if students were allowed to pick larger packages. 
Envy-freeness or stability has been raised as one of the arguments why the Gale-Shapley mechanism for simple assignment problems where agents have unit-demand (i.e. demand for only one course seat) is so successful in practice \citep{Roth2002}. If the market outcome is unstable, there is an agent or apir of agents who have the incentive to circumvent the match. We argue that this property is as important for the assignment of course schedules.
So, it envy-freeness matters, the elegant BPS mechanism has a number of attractive economic properties and is computationally tractable. %much less expensive compared to A-CEEI.

\bibliographystyle{ormsv080}
\bibliography{matching_bundle}
\newpage

\begin{APPENDICES}

\section{Proof}
\label{sec:proof}
\begin{proof}
Since students are interested in seats in more than one course, the sum of capacities of all selectable courses (tutor groups) is	significantly higher than the number of participating students, therefore $\min{Q,|S|} = |S|$

For matching problems with single unit demand, the number of possible ranks equals the number of courses, i.e. $|C| = R$. 	That is, we can rewrite $TA(M) = R\cdot |S|$. Since the students do not rank single courses but bundles of courses, we have to replace $c\in C$ by $b\in B$. We use this to get our conclusion:
\begin{eqnarray*}
	AUPCR\rbr{M} &=& \frac{AUPC\rbr{M}}{TA\rbr{M}} = \frac{\sum_{r=1}^{R} \abs{\{\rbr{i, b} \in M \mid \rank{i}{b} \leq r\}}}{R|S|}\\
	 &=&\frac{1}{R} \sum_{r=1}^{R} \frac{\abs{\{\rbr{i, b} \in M \mid \rank{i}{b} \leq r\}}}{|S|}
\end{eqnarray*}
\end{proof}

\section{Sample Preferences in REV}
\label{app:sample_REV}

It is interesting to understand whether a simple course-level scoring rule as used in \cite{Budish2017} is expressive enough to describe the preference profiles with timely preferences in course assignment. 
Table \ref{tab:REV_param} shows the parameters, the number of generated bundles as well as the respective objective function value of REV for the different sample preferences. None of the preference profiles we tested allowed for a feasible solution in REV.  

\begin{table}
\resizebox{\linewidth}{!}{
\begin{tabular}{|l|c|c|c|c|}\hline
						&student 1  & student 2 &student 3 &student 4\\ \hline
%courses \& gap time		&\multicolumn{4}{c|}{all courses of WS17/18, 15min}\\\hline
min lunch time		 	& \multicolumn{2}{c|}{45 min}&\multicolumn{2}{c|}{0 min}\\\hline
time ranges				& \multicolumn{2}{c|}{8am to 6pm}&\multicolumn{2}{c|}{10am to 6pm}\\\hline
feasible days (score)	& Mo(2), Tu(4), We(5), Th(4), Fr(1)
						& Tu(5), We(5), Th(2)
						& Mo(5), Tu(3), We(5), Th(3)
						& Mo(5), Tu(5), We(5), Th(2), Fr(1)\\\hline
\# of bundles			& $8503$ & $4120$ & $4425$ & $12370$\\\hline
$err(\varepsilon)$		& $\approx 0.005$ & $\approx 0.004$ & $\approx 0.004$ & $\approx 0.005$\\\hline
\end{tabular}}
\caption{Parameters of the REV on the data for the winter term.}
\label{tab:REV_param}
\end{table}

\end{APPENDICES}
\end{document}